\documentclass[12pt,letterpaper]{article}
\usepackage{epsfig,rotating,setspace,latexsym,amsmath,epsf,amssymb,amsfonts,bm,theorem,cite,enumerate,longtable,accents,float,physics,algorithm,graphicx,epsf,authblk,epstopdf,url,xcolor,soul,multirow,bbm,mathrsfs,subcaption,mathtools,comment,dsfont,hyperref,cleveref,booktabs,tree-dvips}
\usepackage[center]{qtree}
\usepackage[linguistics]{forest }
\usepackage[noend]{algpseudocode}

\newtheorem{theorem}{Theorem}

\newtheorem{definition}{Definition}
\newtheorem{remark}{Remark}
\newtheorem{lemma}{Lemma}
\newtheorem{example}{Example}
\newenvironment{Proof}[1]{\medskip\par\noindent{\bf Proof:\,}\,#1}{{\mbox{\,$\blacksquare$}\par}}

\newcommand{\cq}{{\mathcal{Q}}}
\newcommand{\cR}{{\mathcal{R}}}
\newcommand{\cw}{{\mathcal{W}}}
\newcommand{\cs}{{\mathcal{S}}}
\newcommand{\ck}{{\mathcal{K}}}

\newcommand{\cn}{{\mathcal{N}}}

\newcommand{\defeq}{\stackrel{\Delta}{=}}

\DeclareMathOperator{\lcm}{\text{lcm}}

\allowdisplaybreaks

\date{}

\title{The Role of Common Randomness Replication in Symmetric PIR on Graph-Based Replicated Systems\thanks{This work was presented in parts at IEEE Globecom 2025, and (to appear) at IEEE ICC 2026.}}

\author{Shreya Meel \qquad Sennur Ulukus\\
\normalsize Department of Electrical and Computer Engineering\\
\normalsize University of Maryland, College Park, MD 20742 \\
\normalsize {\it smeel@umd.edu} \qquad {\it ulukus@umd.edu}}
 
\begin{document}

\maketitle

\vspace*{-1cm}

\begin{abstract}
   In symmetric private information retrieval (SPIR), a user communicates with multiple servers to retrieve from them a message in a database, while not revealing the message index to any individual server (user privacy), and learning no additional information about the database (database privacy). We study the problem of SPIR on graph-replicated database systems, where each node of the graph represents a server and each link represents a message. Each message is replicated at exactly two servers; those at which the link representing the message is incident. To ensure database privacy, the servers share a set of common randomness, independent of the database and the user's desired message index. We study two cases of common randomness distribution to the servers: i) graph-replicated common randomness, and ii) fully-replicated common randomness. Given a graph-replicated database system, in i), we assign one randomness variable independently to every pair of servers sharing a message, while in ii), we assign an identical set of randomness variable to all servers, irrespective of the underlying graph. In both settings, our goal is to characterize the SPIR capacity, i.e., the maximum number of desired message symbols retrieved per downloaded symbol, and quantify the minimum amount of common randomness required to achieve the capacity. To this goal, in setting i), we derive a general lower bound on the SPIR capacity, and show it to be tight for path and regular graphs through a matching converse. Moreover, we establish that the minimum size of common randomness required for SPIR is equal to the message size. In setting ii), the SPIR capacity improves over the first, more restrictive setting. We show this through capacity lower bounds for a class of graphs, by constructing SPIR schemes from PIR schemes. This includes the class of path and cyclic graphs, for which we derive nearly matching upper bounds, and settle the capacity for the path graph with 3 vertices. Finally,  we extend our results to the $r$-multigraph version of graphs, where every vertex pair with an incident edge shares a set of $r$ messages, and derive bounds on the SPIR metrics. We show that the SPIR capacity bounds are independent of the multiplicity $r$ in the graph-replicated randomness setting, and decrease with $r$ in the fully-replicated randomness setting.
\end{abstract}

\section{Introduction}
Private information retrieval (PIR) \cite{chor} is a cryptographic primitive that safeguards the privacy of a user attempting to retrieve a message in a database, without revealing which message is retrieved. Over the past decade, PIR has generated renewed interest with one of its major focus being the characterization of the capacity \cite{rashmi_erasure_pir}, i.e., the maximum number of desired message symbols retrieved per downloaded symbol. The PIR capacity, for the canonical setting of fully-replicated databases was established by Sun and Jafar \cite{SJ17}, and was optimized for minimum message length by Tian, Sun and Chen \cite{ChaoTian}. The PIR capacity was studied in more realistic settings, such as, PIR with colluding servers \cite{colluding,  arbitrarycollusion, csa},  coded storage \cite{coded_colluding_2017,  Salim_CodedPIR,  mdstpir, banawan_pir_mdscoded,  mds_ge}, eavesdroppers and Byzantine servers \cite{sun_eaves, byzantine_tpir},  single server \cite{kadhe_singleserver_pir, tpir_sideinfo, single-serverSI}, constrained server storage \cite{uncoded_constrainedstorage_pir,  batuhan_hetero},  and graph-replicated storage \cite{graphbased_pir,  BU19, asymp_gxstpir}; we refer the readers to \cite{ulukusPIRLC} and the references therein for more variants of PIR and its applications in private computation and learning. In PIR, user privacy is provided at the cost of revealing parts of the database to the user. This is detrimental in applications where the database contains sensitive information and the privacy of its contents should be preserved. As a solution, symmetric PIR (SPIR) was formulated by Gertner et al. \cite{spir_first} with the additional requirement of database privacy, where no information beyond the desired message is revealed to the user. 

SPIR is not feasible unless some private \emph{common randomness}, independent of the messages, is shared among the servers \cite{spir_first}. The optimal trade-off between SPIR capacity and the minimum amount of common randomness was characterized in \cite{c_spir} in the canonical fully-replicated database setting. Other variants of SPIR were explored in follow-up works, such as SPIR with colluding servers, eavesdroppers, unresponsive and Byzantine servers \cite{tspir_mdscoded, pir_spir_adversaries, nomeir_asymp_bspir}, SPIR on MDS coded messages \cite{skoglund_mds_spir, sun_spir_mds_mismatched}, SPIR with side information \cite{zhusheng_spir_pir, chou_spir}, SPIR with controlled privacy leakage \cite{samy_spir, tian_leakage_pir} and SPIR to retrieve a random message \cite{zhusheng_rspir}. These works assume the availability of all messages and randomness across servers in a coded or an uncoded form. However, the full replication of sensitive information can be limited due to security constraints, e.g., only specific servers are trusted to store certain messages. Moreover, even in fully-replicated database systems, only a subset of messages may be accessible to a user, owing to which the effective storage is not fully-eplicated. For instance, in attribute-based private authentication systems, \cite{ali_dapac, meel_hetdapac} the accessible messages present a database system, whose replication pattern is modeled by a non-uniform hypergraph. This motivates us to study SPIR on non fully-replicated databases, particularly where every message in the database is replicated on two distinct servers.

In this work, we adopt the graph-based storage architecture of the PIR counterpart \cite{graphbased_pir}; here, vertices represent servers, and every message representing an edge between two vertices, is replicated on the two servers corresponding to these vertices. Our system model naturally generalizes to multigraph-based replicated systems, where $r$ messages are replicated between the respective server pairs. We consider two replication models for the server-side common randomness. The first model assumes that every pair of servers that share a set of messages, also share an independent common randomness variable. This may arise when the data distributor assigns a randomness per replicated message only to the designated servers. We refer to this as the \emph{graph-replicated randomness} setting. Next, we relax the constrained randomness availability to the setting where the data distributor can populate common randomness across all servers. We refer to this as the \emph{fully-replicated randomness} setting. The randomness distribution in the first setting enforces stricter privacy constraint, since the randomness is now associated with the messages through shared graph-based replication. Note that, in the canonical setting of \cite{c_spir}, there is no distinction between (hyper)graph-replicated and fully-replicated common randomness settings. 

We establish bounds on the SPIR capacity and minimum common randomness size for various families of graph- and multigraph-based replicated systems. In the graph-replicated common randomness setting, we show that regular (e.g., cyclic, complete)  graphs and path graphs $\mathbb{P}_N$ on $N$ vertices have the same SPIR capacity of $\frac{1}{N}$. For general graphs, the capacity is at least $\frac{1}{N}$, and the minimum required randomness size is $1$ relative to the message size. Further, we show that the SPIR metrics remain unaffected upon increasing the number of messages to $r$ under the multigraph-based replication model. The fully-replicated common randomness setting improves the capacity, which we derive exactly for $\mathbb{P}_3$ to be $\frac{1}{2}$. We provide capacity lower bounds and randomness size upper bounds for path, cyclic, complete and star graphs, by deriving achievable SPIR schemes from the respective PIR schemes. For path and cyclic graphs, we present non-trivial upper bounds in the sense that they are strictly lower than the respective PIR capacities. Unlike the graph-replicated randomness setting, and similar to PIR, the capacity depends on the individual graph structure, and degrades upon extension to multigraph-based replication model. 

The rest of the paper is organized as follows. In Section~\ref{sec:sysmod_spir}, we formally introduce the SPIR problem for simple graphs under the graph-replicated and fully-replicated randomness settings, and state the corresponding results in Section~\ref{sec:main results}. This is followed by the derivation of these results in Sections~\ref{sec:graph-replicated simple} and \ref{sec:fully-replicated simple}. In Section~\ref{sec:ext to multigraphs}, we extrapolate our formulation to uniform multigraphs and present the results under this generalization. Section~\ref{sec:conclude_spir} concludes the paper with directions of future research. 

\begin{figure}[t]
    \centering
    \begin{subfigure}{0.8\textwidth}
    \centering
    \includegraphics[width=0.9\textwidth]{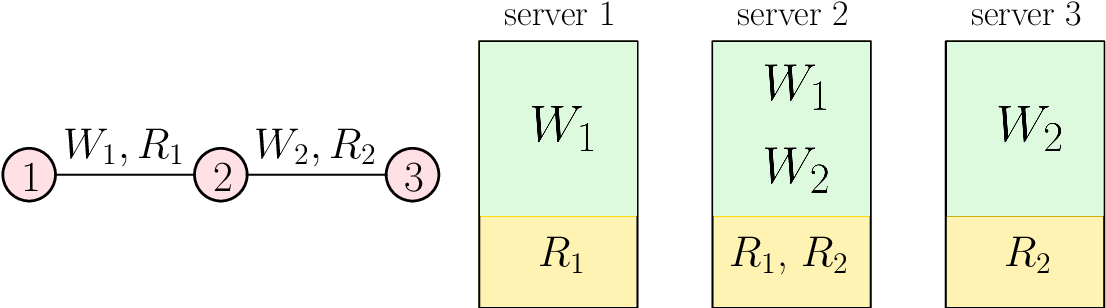} 
    \caption{$K=2$ messages stored on $\mathbb{P}_3$.}
    \label{fig:spir_p3}
    \end{subfigure}
    \begin{subfigure}{0.8\textwidth}
    \centering
    \vspace{0.5cm}
    \includegraphics[width=0.9\textwidth]{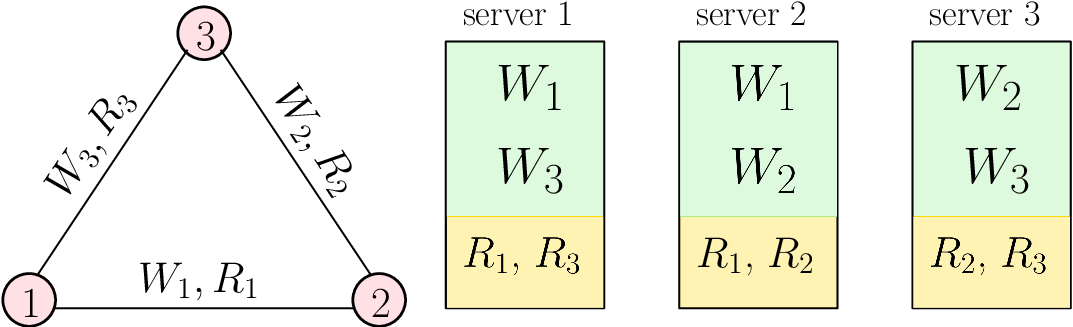}
    \caption{$K=3$ messages stored on $\mathbb{C}_3$.}
    \label{fig:spir_c3}
    \end{subfigure}
    \caption{System model with graph-replicated randomness for $N=3$ servers.}
    \label{fig:spir_path_cycle}
\end{figure}

\section{System Model}\label{sec:sysmod_spir}
In this section, we describe the SPIR problem for simple graphs, i.e., no self-loops and no multi-edges. Consider a database of $K\geq2$ independent messages $\mathcal{W}\defeq\{W_1,\ldots,W_K\}$. Each message $W_k$ comprises $L$ symbols, picked independently and uniformly at random from a finite field $\mathbb{F}_q$, i.e.,
\begin{align}
    H(\cw)=H(W_1)+\ldots+H(W_K)=KL,
\end{align}
where the equality is in $q$-ary units. The messages are stored on $N\geq 2$ non-colluding servers, denoted by the set $\mathcal{S}=[N]\defeq \{1,\ldots,N\}$ in a two-replicated fashion. That is, each message $W_k\in \cw$ is replicated exactly twice and stored on two distinct servers in $\cs$. This storage system is represented by a simple, undirected graph where each vertex represents a server and each edge represents a message. More specifically, in a simple graph $G=(V,E)$, an edge in $E$ is present between two vertices in $V$ if and only if a message in $\cw$ is replicated on the two corresponding servers in $\cs$. Equivalently, $V=\cs$ and $E=\cw$. We assume that $G$ is connected, i.e., there exists a path between each pair of vertices. We consider the following two cases of server-side common randomness replication.

\begin{figure}[t]
    \centering
    \begin{subfigure}{0.85\textwidth}
    \centering
    \includegraphics[width=0.8\textwidth]{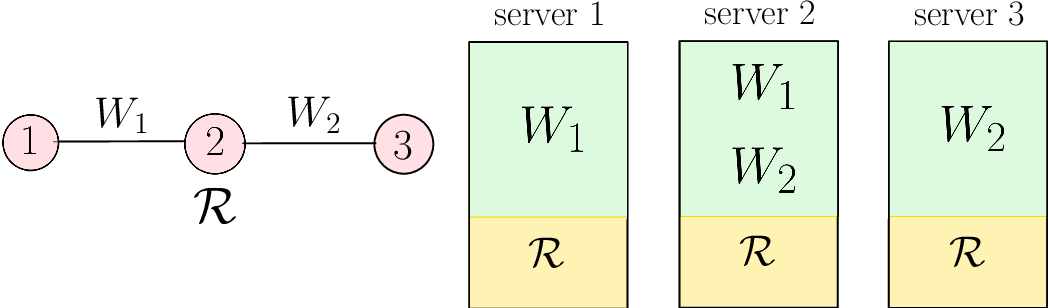} 
    \caption{$K=2$ messages stored on $\mathbb{P}_3$.}
    \label{fig:spir_p3_fullyrep}
    \end{subfigure}
    \begin{subfigure}{0.8\textwidth}
    \centering
    \vspace{0.5cm}
    \includegraphics[width=0.86\textwidth]{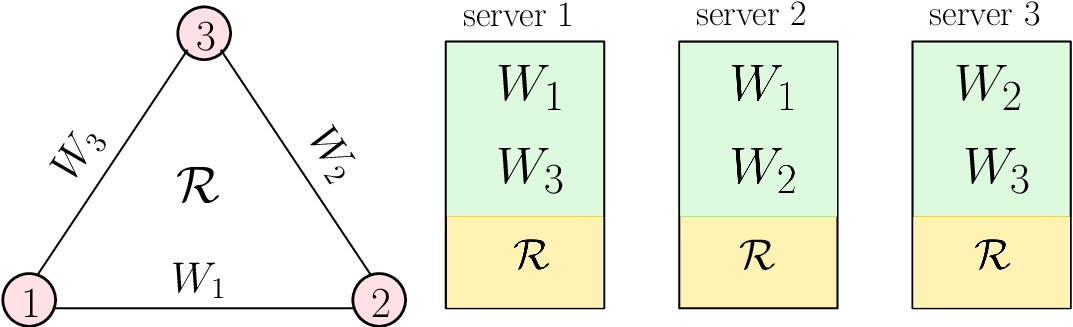}
    \caption{$K=3$ messages stored on $\mathbb{C}_3$.}
    \label{fig:spir_c3_full_rep}
    \end{subfigure}
    \caption{System model with fully-replicated randomness for $N=3$ servers.}
    \label{fig:sysmod_fr}
\end{figure}

\begin{enumerate}[1)]
    \item \textit{Graph-Replicated Common Randomness:}
    In this scenario, for each message $k\in [K]$ we endow the servers sharing $W_k$ with a private random variable $R_k$, independent of $W_k$ and whose realization is unavailable to the user and to the servers that do not store $W_k$. We call this \emph{message-specific common randomness}, where each $R_k$ is also two-replicated. The randomness variables $R_k$ are independent and identically distributed (i.i.d.). Consequently, the total common randomness $\mathcal{R} \defeq\{R_1,\ldots,R_K\}$, replicated according to the same $G$ satisfies,
    \begin{align}
        H(\cR)&=H(R_1)+\ldots+H(R_K) = KH(R_1).
    \end{align}
    This setting is depicted for path and cyclic graphs in Fig.~\ref{fig:spir_path_cycle}.
    
    \item \textit{Fully-Replicated Common Randomness:}
    In this scenario, the total common randomness $\cR$ is available to all servers, independent of $G$, as in the canonical fully-replicated SPIR \cite{c_spir}. This setting is depicted for path and cyclic graphs in Fig.~\ref{fig:sysmod_fr}, 
\end{enumerate}

Let $k$ denote the desired message index, hence, let $W_{k}$ denote the desired message. For both settings, let $\cq$ represent the randomness in the schemes followed by the user to retrieve the messages in $\cw$. Since $\cq$ is decided prior to choosing the message index, it is independent of $k$. Further, $k$ and $\mathcal{Q}$ are independent of $\mathcal{W}$ and $\mathcal{R}$ since the user has no knowledge of the content stored at the servers. To retrieve $W_k$, the user generates $N$ queries $Q_1^{[k]}, \ldots, Q_N^{[k]}$, one for each server, using $\cq$, i.e.,
\begin{align}\label{eq:query_randomness}
    H(Q_1^{[k]},\ldots,Q_N^{[k]}|\cq) = 0,
\end{align}
and sends $Q_i^{[k]}$ to server $i$. Upon receiving the query, server $i$ responds with an answer $A_i^{[k]}$. Let $\mathcal{W}_i$ and $\mathcal{R}_i$ denote the set of messages and randomness stored at server $i$ (note $\cR_i=\cR$ in the second setting). Then, $A_i^{[k]}$ is a deterministic function of $Q_i^{[k]}$, $\mathcal{W}_i$ and $\mathcal{R}_i$, i.e.,
\begin{align}\label{eq:answer_deterministic}
    H(A_i^{[k]}|Q_i^{[k]}, \mathcal{W}_i, \mathcal{R}_i) = 0.
\end{align}

A valid SPIR scheme should meet the following requirements: user privacy, reliability, and database privacy. For user privacy, the query and answer for each server should be identically distributed, irrespective of the desired message index, i.e., for every server $i\in [N]$ and any message index $k\in [K]$,
\begin{align}\label{eq:user_privacy}
    (Q_i^{[k]}, A_i^{[k]},\mathcal{W}_i, \mathcal{R}_i) \sim (Q_i^{[1]}, A_i^{[1]},\mathcal{W}_i, \mathcal{R}_i).
\end{align}
To guarantee reliability, the user should be able to exactly recover their requested message $W_{k}$, using the answers from all the servers, i.e.,
\begin{align}
    H(W_k|A_1^{[k]},\ldots,A_N^{[k]}, \cq) = 0.\label{eq:reliability}
\end{align}
The database privacy requirements are different for both settings, as described next. In the first setting, by the independence of message-specific randomness symbols, for any $k$ and any subset $\mathcal{J}\subseteq [K]\setminus \{k\}$, we require that,
\begin{align}
    I( W_{\mathcal{J}};A_{1:N}^{[k]},Q^{[k]}_{1:N},\cR\setminus R_{\mathcal{J}},\cw\setminus \{W_k,W_{\mathcal{J}}\},\cq)=0,\label{eq:database_privacy1}
\end{align}
where $W_{\mathcal{J}}=\{W_\ell:\ell\in \mathcal{J}\}$, $R_{\mathcal{J}}=\{R_\ell:\ell\in \mathcal{J}\}$, $Q_{1:N}^{[k]}=\{Q_1^{[k]},\ldots,Q_N^{[k]}\}$ and $A_{1:N}^{[k]}=\{A_1^{[k]},\ldots,A_N^{[k]}\}$. In the second setting, the database privacy is defined identically to the canonical definition in \cite{c_spir}, i.e.,
\begin{align}\label{eq:database_privacy2}
    I(W_{\overline{k}};A_1^{[k]},\ldots,A_N^{[k]}|\cq)=0.
\end{align}

\begin{remark}
With $\mathcal{J}=[K]\setminus \{k\}$ in \eqref{eq:database_privacy1}, we obtain $
I({W}_{\overline{k}};A^{[k]}_{1},\ldots,A_N^{[k]},Q^{[k]}_1,\ldots,Q_N^{[k]},R_k,\cq)=0$, which means that the answers from the servers given the queries, reveal no information on ${W}_{\overline{k}}:=\mathcal{W}\setminus W_k$ to the user, even if $R_k$ is retrieved in the process. This differs from \eqref{eq:database_privacy2} and the definition in \cite{c_spir}, due to integration of message-specific common randomness.
\end{remark}

An SPIR scheme under the graph-replicated or fully-replicated randomness setting, is said to be achievable if it simultaneously satisfies \eqref{eq:user_privacy}, \eqref{eq:reliability}, and \eqref{eq:database_privacy1} or \eqref{eq:database_privacy2}, respectively. To quantify the efficiency of an SPIR scheme, we adopt the following metrics. 

\textit{Capacity:} We express the rate of an SPIR scheme on $G$ as the ratio between the number of desired message symbols and the total number of downloaded symbols, i.e.,
\begin{align}
    \text{rate}= \frac{L}{\sum_{i=1}^N H(A_i^{[k]})}.
\end{align}
The capacity of SPIR on $G$ is defined as the supremum over all achievable SPIR rates. For the graph-replicated randomness setting, we use $R(G)$ and $\mathscr{C}(G)$, to denote rate and capacity, respectively. Under the fully-replicated randomness setting, we use the notations $R_{FR}(G)$ and $\mathscr{C}_{FR}(G)$, respectively, where the subscript $FR$ indicates \emph{fully-replicated} randomness. Note that, replicating $\cR$ in all servers leads to better interference alignment in an SPIR scheme, implying savings in the download cost. Thus, for any graph $G$, the capacity $\mathscr{C}_{FR}(G)$ is greater than $\mathscr{C}(G)$, and the following inequality 
\begin{align}\label{relation_capacities}
   \mathscr{C}(G)<\mathscr{C}_{FR}(G)<\mathscr{C}_{PIR}(G)
\end{align}
holds, where $\mathscr{C}_{PIR}(G)$ denotes the capacity of PIR on $G$, i.e., without a database privacy constraint. This implies that $\mathscr{C}_{PIR}(G)$ is a trivial upper bound on $\mathscr{C}_{FR}(G)$, which is a trivial upper bound on $\mathscr{C}(G)$. 

The following metrics quantify the minimum amount of common randomness required for the feasibility of SPIR on $G$. 

\textit{Minimum Randomness Ratio:} In the graph-replicated setting, we measure the common randomness required per message. Towards this, we define \emph{randomness ratio},
\begin{align}
    \rho \triangleq \frac{H(R_k)}{L}, \quad k\in [K],
\end{align} 
as the size of message-specific randomness relative to the size of a message, required for an achievable SPIR scheme. The optimum randomness ratio $\rho^*$ is the minimum value of $\rho$ over all feasible SPIR schemes.

\textit{Minimum Total Randomness Ratio:} In the fully-replicated randomness setting, we define the \emph{total randomness ratio},
\begin{align}
        \rho_{total}(G) \triangleq \frac{H(\cR)}{L},
\end{align}
as the size of total common randomness $\mathcal{R}$, for an achievable SPIR scheme. Note that this definition is redundant for the first setting since $\rho_{total}(G)=K\rho$, due to the independence of $R_k$ for all $k\in [K]$. The optimum total randomness ratio $\rho^*_{total}(G)$ is the minimum value of $\rho_{total}(G)$ over all feasible SPIR schemes on $G$. 

As our main contribution, we characterize bounds on the aforementioned quantities, and show their optimality for certain families of graphs.

\section{Main Results}\label{sec:main results}
In this section, we state the SPIR results for simple graphs $G$ for the graph-replicated and fully-replicated common randomness settings. Note that, if $N=2$, the only connected simple graph is $\mathbb{P}_2$. That is, a single message is replicated on two servers, and both the SPIR settings coincide. Moreover, the PIR and SPIR problems become trivial and the capacity is $1$. We hereby focus on graphs with $N\geq 3$. We begin with the results for the first setting. 

\begin{theorem}\label{thm:achievable_rate_all_G}
    For any graph $G$ with $N$ vertices, its SPIR capacity $\mathscr{C}(G)$ is bounded as
    \begin{align}
        \mathscr{C}(G)\geq \frac{1}{N},
    \end{align}
provided that the randomness ratio $\rho=1$, i.e., $\rho^*\leq\rho = 1$.
\end{theorem} 

\begin{theorem}\label{thm:bound_rho}
For any SPIR scheme on $G$, the required randomness ratio $\rho$ is at least $1$; otherwise SPIR is not feasible.
\end{theorem}

Combining Theorems~\ref{thm:achievable_rate_all_G} and \ref{thm:bound_rho}, we get that $\rho^*=1$ for any $G$. Next, we state the capacity result for specific graphs. In a $d$-regular graph, the degree of every vertex is $d$.

\begin{theorem}\label{thm:upperbnd_capacity}
    If $G=\mathbb{P}_N$ or $G$ is a $d$-regular graph, 
    \begin{align}
        \mathscr{C}(G)\leq \frac {1}{N}.
    \end{align}
\end{theorem}

Theorems \ref{thm:achievable_rate_all_G} and \ref{thm:upperbnd_capacity} imply that $\mathscr{C}(G)=\frac{1}{N}$ for these graphs. The following remarks compare the SPIR capacities of certain graphs with the corresponding PIR capacities.

\begin{remark}
    The PIR capacity for $\mathbb{P}_N$ is $\frac{2}{N}$ \cite{our_journal2025}. Thus, incorporating the database privacy constraint \eqref{eq:database_privacy1} hurts the capacity by exactly half. The PIR capacity for $\mathbb{C}_N$ is $\frac{2}{N+1}$ \cite{BU19}. Since $\mathbb{C}_N$ is a $2$-regular graph, the corresponding SPIR capacity is $\frac{1}{N}$, which is greater than half of its PIR capacity. In general, the PIR capacity for regular graphs with $N$ vertices is atmost $\frac{2}{N}$ \cite{SGT23}. The corresponding SPIR capacity is at least half the PIR capacity.
\end{remark}

\begin{remark}
    We observe a distinct feature of the bounds on SPIR capacity, compared to those on the PIR capacity, in case of regular graphs. Regular graphs with equal $N$ and varying $K$ have equal SPIR capacities, which is not necessarily true for PIR. For instance, consider the cyclic graph $\mathbb{C}_N(d=2)$ with $K=N$, the complete graph $\mathbb{K}_{N}(d=N-1)$ and the complete bipartite graph $\mathbb{K}_{N/2,N/2}(d=N/2)$ for $N$ even, as examples. All of them have $\mathscr{C}(G)=\frac{1}{N}$. The first has PIR capacity $\frac{2}{N+1}$, while for the latter cases, the PIR capacity bounds are strictly distinct \cite{gePIR} from $\frac{2}{N+1}$.
\end{remark}

Before stating the next result, we require the following definition. 

\begin{definition}[Symmetric Retrieval Property (SRP)\cite{our_journal2025}] \label{def:srp}
    A graph-based PIR scheme is said to satisfy the symmetric retrieval property if, for any $k$, the number of symbols of $W_k$ retrieved from each of the servers storing $W_k$, is equal. Equivalently, $H(A_{i}^{[k]}|Q_{i}^{[k]},\cw_{i}\setminus\{W_k\})=H(A_{j}^{[k]}|Q_{j}^{[k]},\cw_{j}\setminus\{W_k\})=\frac{H(W_k)}{2}$, if $W_k$ is replicated on servers $i$ and $j$.
\end{definition}

\begin{theorem}\label{thm:achievable_rate_SRP_G}
    Consider a PIR scheme $T$ on $G$, which satisfies the SRP. Let the resulting PIR rate be $R_{PIR}(G)$. Then, there exists an SPIR scheme for $G$, with rate $R(G)=\frac{R_{PIR}(G)}{2}$ and $\rho = 1$.
\end{theorem}

\begin{remark}
    For the case of $\mathbb{P}_N$, the alternative scheme of Theorem~\ref{thm:achievable_rate_SRP_G} is also capacity-achieving. However, for regular graphs, for which the PIR capacity is less than $\frac{2}{N}$ (e.g., $\mathbb{C}_N, \mathbb{K}_N, \mathbb{S}_N$), this scheme operates strictly below the SPIR capacity.
\end{remark}

\begin{table}[tbp]
    \centering
    \renewcommand{\arraystretch}{1.3}
    \setlength{\tabcolsep}{8pt}
    \begin{tabular}{@{}lccc@{}}
    \toprule
    \textbf{Graph} & \textbf{PIR scheme} & \textbf{General SPIR scheme} & \textbf{PIR-derived SPIR scheme}\\
    \midrule
     Path $\mathbb{P}_N$ & $\color{red}{{\frac{2}{N}}}$ \cite{our_journal2025} & $\color{red}{\frac{1}{N}}$ & $\color{red}{{\frac{1}{N}}}$\\
    
     Cyclic $\mathbb{C}_N$  & $\color{red}{\frac{2}{N+1}}$ \cite{BU19} & $\color{red}{{\frac{1}{N}}}$ & $\frac{1}{N+1}$\\
    
     Complete $\mathbb{K}_N$  & $\geq \left(\frac{4}{3}-o(1)\right)\frac{1}{N}$ \cite{gePIR} & $\color{red}{\frac{1}{N}}$ & $\geq \left(\frac{2}{3}-o(1)\right)\frac{1}{N}$\\
     Star $\mathbb{S}_N$ & $\frac{2}{N}$  \cite{YJ23} & $\frac{1}{N}$ & $\frac{1}{N}$\\
     \bottomrule
    \end{tabular}
    \caption{PIR and SPIR rates with graph-replicated common randomness (capacities highlighted in {\color{red} red}) on some graphs for which PIR schemes satisfying SRP are known.}
    \label{tab:results gr}
\end{table}

The results for the graph-replicated common randomness setting are summarized in Table~\ref{tab:results gr}. Next, we state the results for the fully-replicated randomness setting.

\begin{theorem}\label{thm:spir scheme fr}
    Given a PIR scheme on $G$ with $N$ vertices and $K$ edges, that satisfies SRP, there exists an SPIR scheme with fully-replicated randomness on $G$. Let $L'$ be the number of symbols per message, and $D'$ be the total number of downloaded symbols of the PIR scheme, then the following SPIR rate,
    \begin{align}\label{eq:rate_G}
        R_{FR}(G)=\frac{L'x}{D'x+Ny}\leq \mathscr{C}_{FR}(G),
    \end{align}
    is achievable, where $x=\frac{\ell}{L'/2}$, $y=\frac{\ell}{N-1}$, $\ell=\lcm(L'/2,N-1)$ and $\lcm$ denotes the least common multiple. The corresponding total randomness ratio is,
    \begin{align}\label{eq:rho_total_G}
        \rho^*_{total}(G)\leq\rho_{total}(G)=\frac{K-1}{2}+\frac{Ny}{L'x}.
    \end{align}
\end{theorem}

\begin{remark}\label{rem:lbnd_capacity cycle, path}
    The capacity-achieving schemes of $\mathbb{P}_N$ \cite{our_journal2025} and $\mathbb{C}
    _N$ \cite{BU19} satisfy the SRP. For these graphs \eqref{eq:rate_G} reduces to,
    \begin{align}
        R_{FR}(\mathbb{P}_N) & = \frac{2}{N+\frac{N}{N-1}}, \qquad R_{FR}(\mathbb{C}_N) = \frac{2}{N+1+\frac{N}{N-1}}\label{eq: rate_path cyclic}.
    \end{align}
    This strictly improves upon $\mathscr{C}(\mathbb{P}_N)=\mathscr{C}(\mathbb{C}_N)=\frac{1}{N}$ of Theorem~\ref{thm:upperbnd_capacity}. In fact, the rates are related to the respective PIR capacities by,
    \begin{align}
        \frac{1}{R_{FR}(G)}=\frac{1}{\mathscr{C}_{PIR}(G)}+\frac{N}{2(N-1)}.
    \end{align}
    Further, \eqref{eq:rho_total_G} simplifies to $\rho_{total}(G)={R_{\text{F-R}}(G)}^{-1}-1$. Consequently, we have,
    \begin{align}
        \rho^*_{total}(\mathbb{P}_N)&\leq \rho_{total}(\mathbb{P}_N)=\frac{N-2}{2}+\frac{N}{2(N-1)},\\
        \rho^*_{total}(\mathbb{C}_N)&\leq \rho_{total}(\mathbb{C}_N)=\frac{N-1}{2}+\frac{N}{2(N-1)}. 
    \end{align}
\end{remark}

\begin{theorem}\label{thm:capacity of P_3 fr}
    For the special case of $\mathbb{P}_3$, we identify the capacity as,
    \begin{align}
        \mathscr{C}_{\text{FR}}(\mathbb{P}_3)=\frac{1}{2}.
    \end{align}
\end{theorem}

\begin{remark}
    Note that Theorem~\ref{thm:spir scheme fr} yields a lower rate $\frac{4}{9}$ on $\mathbb{P}_3$, compared to $\mathscr{C}_{FR}(\mathbb{P}_3)=\frac{1}{2}$. The lower bound on the capacity  is achieved through a simple scheme on star graphs described in Section~\ref{subsec:star scheme}, and notably does not require the underlying PIR scheme to satisfy SRP.     
\end{remark}

\begin{remark}
    Although not capacity-achieving, there is a PIR scheme on $\mathbb{S}_N$ satisfying SRP with $R_{PIR}(\mathbb{S}_N)=\frac{2}{N}$. The SPIR scheme through the algorithm in Section~\ref{subsec:achievable_algo} attains the metrics of $\mathbb{P}_N$. In Section~\ref{subsec:star scheme}, we discuss how PIR schemes on $\mathbb{S}_N$ with best-known rates can be converted to the SPIR schemes.
\end{remark}

The next theorem provides an upper bound on $\mathscr{C}_{FR}(G)$, and a lower bound on $\rho^*(G)$ for path and cyclic graphs.

\begin{theorem}\label{thm:upperbnd_capacity_fr}
    The capacity and minimum total randomness ratio of  path and cyclic graphs satisfy
    \begin{align}
        \mathscr{C}_{FR}(\mathbb{P}_N)&\leq \frac{2}{N+\frac{2}{N-1}},  &\rho^*_{total}(\mathbb{P}_N)&\geq \frac{N-2}{2}+\frac{1}{N-1},\\
        \mathscr{C}_{FR}(\mathbb{C}_N)&\leq \frac{2}{N+1+\frac{1}{N-1}},  &\rho^*_{total}(\mathbb{C}_N)&\geq \frac{N-1}{2}+\frac{1}{2(N-1)}.
    \end{align}
\end{theorem}

Although the bounds of Remark~\ref{rem:lbnd_capacity cycle, path} and Theorem~\ref{thm:upperbnd_capacity_fr} do not match, the latter improves upon the trivial capacity bounds $\mathscr{C}_{PIR}(\mathbb{P}_N)$ and $\mathscr{C}_{PIR}(\mathbb{C}_N)$, respectively.

\section{SPIR with Graph-Replicated Randomness}\label{sec:graph-replicated simple}
In this section, we derive our results on graph-replicated common randomness. We start with the general achievable scheme in Section~\ref{sec:scheme1} which proves Theorem~\ref{thm:achievable_rate_all_G}. Next, we derive the lower bound on $\rho^*$ for the feasibility of SPIR, followed by the matching converse for regular and path graphs. Finally, in Section~\ref{sec:spir from pir gr}, we describe the construction of the SPIR scheme derived from an SRP-satisfying PIR scheme on $G$. 

\subsection{Achievability Proof}\label{sec:scheme1}
The scheme achieving $R(G)=\frac{1}{N}$ at $\rho=1$ is inspired from Raviv et al.'s PIR scheme \cite{graphbased_pir} on any two-replicated system (which yields $R_{PIR}(G)=\frac{1}{N}$), coupled with one-time padding for database privacy. Remarkably, the additional database privacy requirement does not hurt the achievable SPIR rate and the same PIR rate of $\frac{1}{N}$ is achievable. 

\subsubsection{General Scheme for Graphs}\label{subsec:gen_scheme_graphs}
Let $I(G)$ denote the incidence matrix of the graph $G$. That is, given $G$, $I(G)$ is defined as the $|V|\times |E| = N\times K$ binary matrix, where rows represent vertices and columns represent the edges. The $(i,j)$th entry of $I(G)$ is 1 if edge $j$ is incident with vertex $i$ and $0$ otherwise. The degree $\text{deg}(i)$ of vertex $i$ is the number of edges incident with it. For the SPIR system based on $G$, suppose each message consists of a single symbol, i.e., $L=1$. Further, assume that $\rho = 1$, i.e., $H(R_\ell) = L$ for all $\ell\in [K]$. Let $k$ be the desired message index. For each server $n\in [N]$, we arrange the message indices at server $n$ in an ascending order, and denote it by the ordered set $\mathcal{F}_n = (\ell:W_\ell\in \mathcal{W}_n)$. Clearly, $|\mathcal{F}_n| = \text{deg}(n)$. Further, we represent $\mathcal{W}_n$ as a vector $\bm{W}_n = [W_\ell, \ell\in \mathcal{F}_n]^{\top}$.
 
Each column of $I(G)$ has exactly two $1$'s. Let us write the signed incidence matrix $\bar{I}(G)$ by replacing the lower $1$-entry to $-1$ for each column. This matrix is known to the servers, since they know $G$. To privately retrieve $W_k$, the user chooses $K$ symbols $(h_1, h_2, \ldots, h_K)$ independently and uniformly at random, and forms the base query matrix,
\begin{align}
    \bm{H} = \bar{I}(G)\cdot \text{diag}(h_1,\ldots,h_K).
\end{align}
Let $\bm{h}_n$ denote the $n$th row of $\bm{H}$ after discarding the zeros. Then, the user sends the following queries, where $\bm{e}_m$ is the standard unit column vector with $1$ at the $m$th place,
\begin{align}\label{eq:queries_gen}
    Q_n^{[k]} = \begin{cases}
        \bm{h}_n^{\top}, & n \in [N]\setminus \{j\},\\
        \bm{h}_n^{\top} + \bm{e}_m, & n = j,
    \end{cases}
\end{align}
assuming that $W_k$ is replicated at servers $i$ and $j$ and that $\bm{e}_m^{\top}\bm{W}_j=W_k$. Server $n$ responds with the following answer symbol for all $n\in [N]$,
\begin{align}\label{eq:ans_gen}
    A_n^{[k]} = Q_n^{[k]\top}\bm{W}_n + \sum_{\ell\in \mathcal{F}_n}\bar{I}(G)(n,\ell)\cdot R_\ell.
\end{align}

\textit{Reliability:} Note that, by the design of queries and answers, 
\begin{align}
A_n^{[k]} = \begin{cases}
    \sum_{\ell\in \mathcal{F}_n} \bar{I}(G)(n,\ell) (h_\ell W_\ell + R_{\ell}) ,& n\in [N]\setminus \{j\},\\
    \sum_{\ell\in \mathcal{F}_n} \bar{I}(G)(n,\ell) (h_\ell W_\ell + R_{\ell})+W_k, &n = j.   
\end{cases}
\end{align}
 To decode $W_{k}$ for any $k\in [K]$, the user computes the sum of all the answers, which ensures reliability since,
\begin{align}
    \sum_{n\in [N]} A_n^{[k]} &=\sum_{n\in [N]}\bigg( \sum_{\ell\in \mathcal{F}_n} \bar{I}(G)(n,\ell) (h_\ell W_\ell + R_{\ell})\bigg)+W_k\\
    &= \sum_{\ell\in \mathcal{F}_n} (h_\ell W_\ell + R_{\ell}) \bigg( \sum_{n\in [N]} \bar{I}(G) (n,\ell) \bigg) + W_k \label{eq:answer decoding}\\
    &=W_k, \label{eq:row sum of I bar G is 0}
\end{align}
where \eqref{eq:row sum of I bar G is 0} follows because the entries of any column $\ell\in [K]$ of $\bar{I}(G)$ sum to 0, proving the reliability constraint \eqref{eq:reliability}. In fact, each $h_\ell W_\ell$ and $R_\ell$ appear twice with opposite signs in \eqref{eq:answer decoding}.

\textit{User Privacy:} For any desired message index $k$, server $n$ receives a query vector of length $\deg(n)$. Because the servers do not collude, each server observes a uniformly distributed random vector over $\mathbb{F}_q^{\deg(n)}$. To compute the answer, server $n$ performs the same linear computation \eqref{eq:ans_gen}, irrespective of $k$. Hence, the user privacy constraint \eqref{eq:user_privacy} is satisfied. 

\textit{Database Privacy:} On a high level, \eqref{eq:database_privacy1} is satisfied since, besides $W_k$, the user receives linear combinations of $W_\ell,$ masked with $R_\ell$ (with suitable signs), $ \ell\in [K]$. Let $\mathcal{J}= \{\ell_1,\ldots, \ell_{|\mathcal{J}|}\}$ be a subset of $[K]\setminus \{k\}$. Given $\cw\setminus W_{\mathcal{J}}$, $\cR\setminus R_{\mathcal{J}}$ and $\cq$, the answers $A_{1:N}^{[k]}$ can be effectively expressed as the matrix-vector product,
\begin{align}
    \begin{bmatrix}
    \bar{I}(G)[:,\ell_1],& \ldots, &  \bar{I}(G)[:,\ell_{|\mathcal{J}|}]
    \end{bmatrix}\begin{bmatrix}
        h_{\ell_1}\left(W_{\ell_1} + \frac{1}{h_{\ell_1}}R_{\ell_1}\right)\\
        \vdots\\
        h_{\ell_{|\mathcal{J}|}}\left(W_{\ell_{|\mathcal{J}|}} + \frac{1}{h_{\ell_{|\mathcal{J}|}}}R_{\ell_{|\mathcal{J}|}}\right) 
    \end{bmatrix},
\end{align}
where $ \bar{I}(G)[:,\ell]$ is the $\ell$th column of $\bar{I}(G)$, with the matrix  $\begin{bmatrix} \bar{I}(G)[:,\ell_1],& \ldots, &  \bar{I}(G)[:,\ell_{|\mathcal{J}|}] \end{bmatrix}$ and coefficients $h_{\ell_1}, \ldots, h_{\ell_{|\mathcal{J}|}}$ available to the user. By the independence of $R_{\ell_1}, \ldots, R_{\ell_{|\mathcal{J}|}}$, which one-time pad \cite{shannon_otp} the messages, the user learns no information on $W_{\mathcal{J}}$.  

The scheme requires downloading a single $\mathbb{F}_q$ symbol as answer from each of the $N$ servers, to recover $L=1$ symbol of the desired message. This results in the SPIR rate $R(G) = \frac{1}{N}$.
\begin{remark}
    Note that our scheme bears similarity with that of secure summation e.g., \cite{zhao_securesum}, where randomness symbols that sum to 0 are employed to provide information-theoretic security. The decoding step in our scheme is computing the summation of answers; hence, the idea of using zero-sum randomness is common in both schemes.
\end{remark}

\subsubsection{Illustrative Examples}
Next, we illustrate our scheme on several example graphs. The signed incidence matrices $\bar{I}(G)$ for the graphs used in the following three examples are tabulated in Table~\ref{tab:signed incidences}.

\begin{table}[t]
    \centering
    \renewcommand{\arraystretch}{1.2}  
    \setlength{\tabcolsep}{8pt}       
    \begin{tabular}{@{}c| c@{}}        
    \toprule
    $N = 3$ &
    $\displaystyle
    \bar{I}(\mathbb{P}_3) =
    \begin{bmatrix}
      1 & 0 \\
     -1 & 1 \\
      0 & -1
    \end{bmatrix},
    \quad
    \bar{I}(\mathbb{C}_3) =
    \begin{bmatrix}
      1 & 0 & 1 \\
     -1 & 1 & 0 \\
      0 & -1 & -1
    \end{bmatrix}
    $ \\
    \midrule
    $N = 4$ &
    $\displaystyle
    \bar{I}(\mathbb{S}_4) =
    \begin{bmatrix}
      1 & 0 & 0 \\
      0 & 1 & 0 \\
      0 & 0 & 1 \\
     -1 & -1 & -1
    \end{bmatrix},
    \quad
    \bar{I}(\mathbb{M}) =
    \begin{bmatrix}
      1 & 1 & 0 & 0 \\
     -1 & 0 & 1 & 0 \\
      0 & -1 & -1 & 1 \\
      0 & 0 & 0 & -1
    \end{bmatrix}
    $ \\
    \bottomrule
    \end{tabular}
    \caption{Signed incidence matrices for the example graphs.}
    \label{tab:signed incidences}
\end{table}

\begin{example}[Path graph]
Consider the path graph $\mathbb{P}_3$, as shown in Fig.~\ref{fig:spir_path_cycle}(a) where each message and common randomness consist of a single symbol from $\mathbb{F}_q$, i.e., $L=1$. The user chooses two random symbols $h_1$ and $h_2$ from $\mathbb{F}_q$, and forms the matrix,
\begin{align}
    \bm {H}=
    \begin{bmatrix}
        h_1 & 0\\
        -h_1 & h_2\\
        0 & -h_2
    \end{bmatrix}.
\end{align}
Let the desired message index be $\theta$. Then, the queries sent are as follows,
\begin{align}
    Q_1^{[\theta]}=h_1, \qquad Q_2^{[\theta]} = [-h_1, \quad h_2]^\top+\bm{e}_{\theta}, \qquad Q_3^{[\theta]} =-h_2.
\end{align}
The answers returned by the servers are:
\begin{align}
    A_1^{[\theta]}&=h_1W_1+R_1, \\
    A_2^{[\theta]}&= -h_1W_1+h_2W_2+W_{\theta}-R_1+R_2,\\
    A_3^{[\theta]}&= h_2W_2-R_2.
\end{align}
To decode $W_{\theta}$, the user computes the sum of all the answers, resulting in the rate $\frac{1}{3}$.
\end{example}

\begin{example}[Cyclic graph]\label{ex:cyclic graph general scheme}
Consider the cyclic graph $\mathbb{C}_3$ as shown in Fig.~\ref{fig:spir_path_cycle}(b), with $L=1$ and $\rho=1$. The user chooses the random symbols $h_1,h_2,h_3$ from $\mathbb{F}_q$, forming the vectors
\begin{align}
    \bm{h}_1=[h_1, \ h_3]^\top, \qquad \bm{h}_2=[-h_1, \ h_2]^\top, \qquad \bm{h}_3=[-h_2, \ -h_3]^\top.
\end{align}
The answers returned by the servers are shown in Table \ref{tab:scheme1_cycle}, where $\bm{W}_1=[W_1 \ W_3]^\top, \bm{W}_2=[W_1 \ W_2]^\top$ and $\bm{W}_3=[W_2 \ W_3]^\top$. The user adds all the answers to obtain $W_{\theta}$. The scheme results in the rate $\frac{1}{3}$.
\end{example}

\begin{table}[h]
    \centering
    \begin{tabular}{|c|c|c|c|}
    \hline
    & server 1 & server 2 & server 3\\
    \hline
    $\theta=1$ & $\bm{h}_1^\top\bm{W}_1+R_1+R_3$ & $\bm{h}_2^\top\bm{W}_2+W_1-R_1+R_2$ & $\bm{h}_3^\top\bm{W}_3-R_2-R_3$\\
    \hline
    $\theta=2$ & $\bm{h}_1^\top\bm{W}_1+R_1+R_3$ & $\bm{h}_2^\top\bm{W}_2-R_1+R_2$ & $\bm{h}_3^\top\bm{W}_3+W_2-R_2-R_3$\\
    \hline
    $\theta=3$ & $\bm{h}_1^\top\bm{W}_1+W_3+R_1+R_3$ & $\bm{h}_2^\top\bm{W}_2-R_1+R_2$ & $\bm{h}_3^\top\bm{W}_3-R_2-R_3$\\
    \hline
    \end{tabular}
    \caption{Answers for SPIR on $\mathbb{C}_3$.}
    \label{tab:scheme1_cycle}
\end{table}

\begin{example}[Star graph]\label{ex:star graph general scheme}
\begin{figure}[t]
    \centering
    \begin{subfigure}[t!]{0.49\textwidth}
    \centering
    \includegraphics[width=0.5\linewidth]{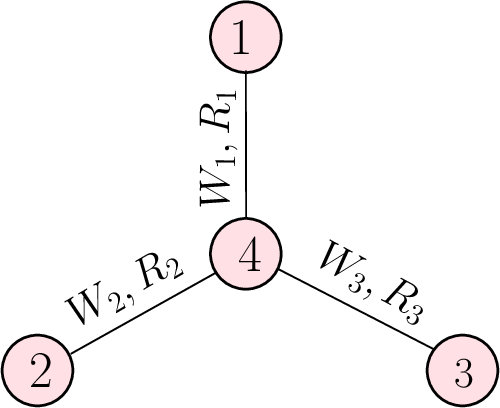}
    \subcaption{Star graph $\mathbb{S}_4$.}
    \vspace{2mm}
    \label{fig:spir_s4}
    \end{subfigure}%
    \hfill
    \begin{subfigure}[t!]{0.49\textwidth}
    \centering
    \includegraphics[width=0.75\linewidth]{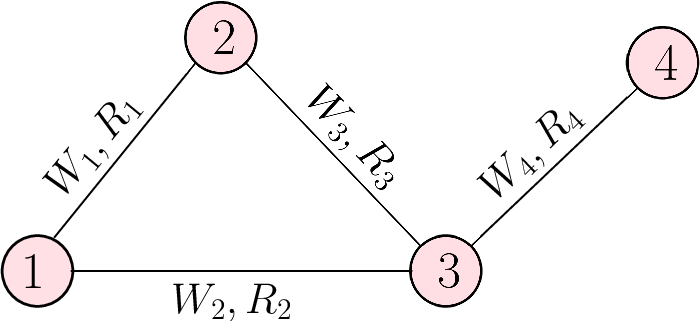}
    \subcaption{Graph $\mathbb{M}$.}
    \label{fig:spir_g4}
    \end{subfigure}
    \caption{SPIR systems with $N=4$ servers.}
     \label{fig:fig3}
\end{figure}

For $N=3$, the star and path graphs are identical. Hence, we demonstrate our achievable scheme for $\mathbb{S}_4$, see Fig.~\ref{fig:fig3}(a). Let server $k$ store $W_k$ and $R_k$ for $k\in [3]$ and server $4$ store all the messages $W_1,W_2,W_3$ and randomness symbols $R_1,R_2,R_3$. To retrieve the desired message $W_{\theta}$, the answers returned are: $h_1W_1+R_1, h_2W_2+R_2, h_3W_3+R_3, -(h_1W_1+h_2W_2+h_3W_3+R_1+R_2+R_3)+W_{\theta}$ by servers $1,2,3$ and $4$, respectively. Clearly, the rate is $\frac{1}{4}$.
\end{example}

\begin{example}
Consider the SPIR system on the graph $\mathbb{M}$ in Fig.~\ref{fig:fig3}(b), with $L=1$ and $\rho=1$. 
The user chooses the random symbols $h_1$, $h_2$, $h_3$ and $h_4$, from $\mathbb{F}_q$ and accordingly sends the queries to the servers using \eqref{eq:queries_gen}. For example, if $\theta=3$, the answers returned are:
\begin{align}
    A_1^{[3]} &= h_1W_1+h_2W_2+R_1+R_2,\\
    A_2^{[3]} &= -h_1W_1+h_3W_3-R_1+R_3,\\
    A_3^{[3]} &= -h_2W_2-h_3W_3+W_3+h_4W_4-R_2-R_3+R_4,\\
    A_4^{[3]} &= -h_4W_4-R_4.
\end{align}
The user decodes $W_3$ by computing the sum of the answers, and the resulting rate is $\frac{1}{4}$.
\end{example}

\subsection{Converse Proofs}\label{sec:converse regular and path}
We make the following observation. Since the scheme is known globally, the user can perform the answer generation of server $n$ and obtain $R_j\in \cR_n$ from the received answer, given $\cw_n$ and $\cR_n\setminus \{R_j\}$, i.e.,
\begin{align}
   H(R_j|A_n^{[k]}, \cw_n,\cR_n\setminus\{R_j\},Q_n^{[k]})=0.\label{eq:R_k_deterministic} 
\end{align} We need the following lemmas. The first lemma is an extension of \cite[Lemma~1]{c_spir} and \cite[Proposition~2]{SGT23} to our setting. 

\begin{lemma}\label{lem:answers_independent_of_index}
    For any subsets $\mathcal{J},\ck\subseteq [K]$, let $W_{\mathcal{J}}=\{W_{\ell}:\ell\in \mathcal{J}\}$ and $R_{\ck}=\{R_{\ell}:\ell\in \ck\}$. Then, for any server $n\in [N]$ and any $k,k'\in [K]$,
    \begin{align}
        H(A_n^{[k]}|W_{\mathcal{J}},R_{\ck}, Q_n^{[k]})=H(A_n^{[k']}|W_{\mathcal{J}},R_{\ck}, Q_n^{[k']}).
    \end{align}
\end{lemma}

The proof of Lemma~\ref{lem:answers_independent_of_index} follows from the user privacy constraint \eqref{eq:user_privacy} of server $n$ and the fact that $A_n^{[k]}$  does not depend on the part of $W_{\mathcal{J}}$ and $R_{\mathcal{K}}$ not intersecting $\mathcal{W}_n$ and $\mathcal{R}_n$, respectively.
Next, the following lemma is an extension of \cite[Lemma~2]{c_spir}.

\begin{lemma}\label{lem:answer_indep_randomness_given_query}
  For any subsets $\mathcal{J},\ck\subseteq [K]$, 
  \begin{align}
      H(A_n^{[k]}|W_{\mathcal{J}},R_{\ck}, Q_n^{[k]})=H(A_n^{[k]}|W_{\mathcal{J}},R_{\ck}, Q_n^{[k]},\cq).
  \end{align}
\end{lemma}

\begin{Proof}
  For this, we show that $I(A_n^{[k]};\cq|W_{\mathcal{J}},R_{\ck}, Q_n^{[k]})=0$, which holds since, 
  \begin{align}
      I(&A_n^{[k]};\cq|W_{\mathcal{J}},R_{\ck}, Q_n^{[k]}) \notag\\
      \leq& I(A_n^{[k]},\cw_{n}\setminus\{W_{\mathcal{J}}\cap\cw_n\},\cR_n\setminus\{R_{\mathcal{K}}\cap\cR_n\};\cq|W_{\mathcal{J}},R_{\ck}, Q_n^{[k]})\\
      =& I(\cw_{n}\setminus\{W_{\mathcal{J}}\cap\cw_n\},\cR_n\setminus\{R_{\mathcal{K}}\cap\cR_n\};\cq|W_{\mathcal{J}},R_{\ck}, Q_n^{[k]})+ I(A_n^{[k]};\cq|\mathcal{W}_n,\cR_n, Q_n^{[k]})\label{eq:last term is zero}\\
      \leq&I(\cw_{n}\setminus\{W_{\mathcal{J}}\cap\cw_n\},\cR_n\setminus\{R_{\mathcal{K}}\cap\cR_n\};\cq|W_{\mathcal{J}},R_{\ck}, Q_n^{[k]}) + I(W_{\mathcal{J}},R_{\ck};\cq|Q_n^{[k]})\label{eq:follows by answer generation}\\
      =&I(\cw_n,\cR_n;\cq|Q_n^{[k]})\\
      \leq& I(\cw_n,\cR_n;\cq,Q_n^{[k]})=0,
  \end{align}
where \eqref{eq:follows by answer generation} follows from \eqref{eq:last term is zero}, since $H(A_n^{[k]}|\cw_n,\cR_n,Q_n^{[k]})=0$.
\end{Proof}

The next lemma is an extension of \cite[Lemma~3]{SGT23} modified for database privacy \eqref{eq:database_privacy1}.

\begin{lemma}\label{lem:sum_entropy of answers} For any two servers $i$ and $j$ that share message $W_k$, it holds that,
    \begin{align}
        H(A_i^{[k]}|\cq)+H(A_j^{[k]}|\cq)&\geq H(A_i^{[k]}|\cR\setminus\{R_k\},W_{\overline{k}},\cq)+H(A_j^{[k]}|\cR\setminus\{R_k\},W_{\overline{k}},\cq)\\
        &\geq (1+\rho)L.
    \end{align} 
\end{lemma} 

\begin{Proof}
    By definition of the lengths of $W_k$ and $R_k$, and their independence from $\cq, W_{\overline{k}}$ and $\cR\setminus\{R_k\}:=R_{\overline{k}}$ we have,
    \begin{align}
        L(1+\rho) =& H(W_k,R_k|\cq)\\
        =&H(W_k|R_{\overline{k}},W_{\overline{k}},\cq)-H(W_k|A^{[k]}_{1:N},R_{\overline{k}},W_{\overline{k}},\cq)+H(R_k)\label{eq:decodability of Wk}\\
        =&H(W_k|R_{\overline{k}},W_{\overline{k}},\cq)-H(W_k|A^{[k]}_i, A^{[k]}_j,R_{\overline{k}},W_{\overline{k}},\cq)+H(R_k)\label{eq:remaining answers deterministic}\\
        =&I(A_i^{[k]},A_j^{[k]};W_k|R_{\overline{k}},W_{\overline{k}},\cq)+H(R_k)\\
        =&H(A_i^{[k]},A_j^{[k]}|R_{\overline{k}},W_{\overline{k}},\cq)-H(A_i^{[k]},A_j^{[k]}|R_{\overline{k}},\cw,\cq)+H(R_k)\\
        =&H(A_i^{[k]},A_j^{[k]}|R_{\overline{k}},W_{\overline{k}},\cq)-I(R_k;A_i^{[k]},A_j^{[k]}|R_{\overline{k}},\cw,\cq)+H(R_k)\label{eq:answers deterministic given all messages and randomness}\\
        =&H(A_i^{[k]},A_j^{[k]}|R_{\overline{k}},W_{\overline{k}},\cq)-H(R_k|R_{\overline{k}},\cw,\cq) + H(R_k|A_i^{[k]},A_j^{[k]},R_{\overline{k}},\cw,\cq)+H(R_k)\\
        = &H(A_i^{[k]},A_j^{[k]}|R_{\overline{k}},W_{\overline{k}},\cq)+H(R_k|A_i^{[k]},A_j^{[k]},R_{\overline{k}},\cw,\cq)\\
        \leq& H(A_i^{[k]},A_j^{[k]}|R_{\overline{k}},W_{\overline{k}},\cq)+H(R_k|A_i^{[k]},R_{\overline{k}},\cw,\cq)\\
        =&H(A_i^{[k]},A_j^{[k]}|R_{\overline{k}},W_{\overline{k}},\cq) \label{eq:R_k_decodability}\\
        \leq& H(A_i^{[k]}|R_{\overline{k}},W_{\overline{k}},\cq)+H(A_j^{[k]}|R_{\overline{k}},W_{\overline{k}},\cq).
    \end{align}
where \eqref{eq:decodability of Wk} follows because $W_k$ can be decoded given all the answers, \eqref{eq:remaining answers deterministic} follows because, except $A_i^{[k]}$ and $A_j^{[k]}$, the answers are functions of $W_{\overline{k}}$ and $\cR\setminus R_k$, and \eqref{eq:answers deterministic given all messages and randomness} follows since $H(A_i^{[k]},A_j^{[k]}|\cR,\cw,\cq)=0$. Finally, \eqref{eq:R_k_decodability} follows from \eqref{eq:R_k_deterministic}.
\end{Proof}

\subsubsection{Proof of Theorem~\ref{thm:bound_rho}}
Let $W_k,R_k$ be stored on servers $i$ and $j$. For the desired message index, $k'\neq k$, from database privacy \eqref{eq:database_privacy1}, choosing $\mathcal{J}=\{k\}$, we have,
\begin{align}
    0=& I(W_k;A_1^{[k']},\ldots,A_N^{[k']},R_{\overline{k}},W_{\overline{k}}\setminus \{W_{k'}\},\cq)\\
    =&  I(W_k;A_1^{[k']},\ldots,A_N^{[k']}|R_{\overline{k}},W_{\overline{k}}\setminus \{W_{k'}\},\cq)+I(W_k;W_{k'}|A_1^{[k']},\ldots,A_N^{[k']},R_{\overline{k}},W_{\overline{k}}\setminus \{W_{k'}\},\cq)\label{eq:decodability of W_k'}\\
    =& I(W_k;W_{k'},A_1^{[k']},\ldots,A_N^{[k']}|R_{\overline{k}},W_{\overline{k}}\setminus \{W_{k'}\},\cq)\\
    =& I(W_k;A_1^{[k']},\ldots,A_N^{[k']}|R_{\overline{k}},W_{\overline{k}},\cq)\label{eq:independence_of_messages}\\
    =& I(W_k;A_i^{[k']},A_j^{[k']}|R_{\overline{k}},W_{\overline{k}},\cq)\label{eq:answers other than i and j exactly deterministic}\\
    \geq & I(W_k;A_i^{[k']}|R_{\overline{k}},W_{\overline{k}},\cq)\\
    =& I(W_k;A_i^{[k]}|R_{\overline{k}},W_{\overline{k}},Q_i^{[k]})\label{eq:consequence of lemma 1 gr}\\
    =&H(A_i^{[k]}|W_{\overline{k}},R_{\overline{k}},Q_i^{[k]})-H(A_i^{[k]}|\cw,\cR,Q_i^{[k]})-I(R_k;A_i^{[k]}|\cw,R_{\overline{k}},Q_i^{[k]})\\
    =&H(A_i^{[k]}|W_{\overline{k}},R_{\overline{k}},Q_i^{[k]})-I(R_k;A_i^{[k]}|\cw,R_{\overline{k}},Q_i^{[k]})\label{eq:Ai_deterministic given everything else}\\
    =& H(A_i^{[k]}|W_{\overline{k}},R_{\overline{k}},Q_i^{[k]})-H(R_k)\label{eq:R_k_deterministic given Ai}\\
    = & H(A_i^{[k]}|W_{\overline{k}},R_{\overline{k}},\cq)-H(R_k), \label{eq:answer i_randomness}
\end{align}
where \eqref{eq:decodability of W_k'} follows since $I(W_k;W_{k'}|A_1^{[k']},\ldots,A_N^{[k']},R_{\overline{k}},W_{\overline{k}}\setminus \{W_{k'}\},\cq)=0$ by \eqref{eq:reliability}, \eqref{eq:independence_of_messages} follows by the independence of messages, \eqref{eq:consequence of lemma 1 gr} is a consequence of \eqref{eq:query_randomness}, Lemma \ref{lem:answers_independent_of_index} and Lemma \ref{lem:answer_indep_randomness_given_query}, \eqref{eq:R_k_deterministic given Ai} is due to \eqref{eq:R_k_deterministic} and \eqref{eq:answer i_randomness} is due to Lemma \ref{lem:answer_indep_randomness_given_query} and \eqref{eq:query_randomness}. 
Similarly, we have,
\begin{align}\label{eq:answer j_randomness}
    H(A_j^{[k]}|W_{\overline{k}},R_{\overline{k}},\cq)-H(R_k)\leq0.
\end{align}
Adding \eqref{eq:answer i_randomness} and \eqref{eq:answer j_randomness}, we get $2H(R_k)\geq H(A_i^{[k]}|W_{\overline{k}},R_{\overline{k}},\cq) + H(A_j^{[k]}|W_{\overline{k}},R_{\overline{k}},\cq)\geq (1+\rho)L$ by Lemma~\ref{lem:sum_entropy of answers}. Substituting $H(R_k)=\rho L$, we obtain $\rho\geq 1$.

\subsubsection{Proof of Theorem~\ref{thm:upperbnd_capacity}}
We derive the exact SPIR capacity of $d$-regular and path graphs. To respect user privacy \eqref{eq:user_privacy}, the result of Lemma \ref{lem:sum_entropy of answers}, combined with Theorem \ref{thm:bound_rho},
\begin{align}\label{eq:pair_ans_bound}
    H(A_i^{[k]}|\cq)+H(A_j^{[k]}|\cq)\geq 2L
\end{align} 
should hold for every pair of servers $\{i,j\}$ which share a file, for all message indices $k$. 

\textit{$d$-regular graph $G$:} By taking the sum of \eqref{eq:pair_ans_bound} over all $\{i,j\}\in E$,
\begin{align}
    \sum_{\{i,j\}\in E}  H(A_i^{[k]}|\cq)+H(A_j^{[k]}|\cq)&\geq 2KL 
\end{align}
which, because $G$ is $d$-regular yields,
\begin{align}
    & d\left(\sum_{n=1}^N H(A_n^{[k]}|\cq)\right) \geq 2KL.
\end{align}
Then, by $Nd=2K$, this results in, 
\begin{align}
    \frac{L}{\sum_{n=1}^N H(A_n^{[k]})}\leq  \frac{L}{\sum_{n=1}^N H(A_n^{[k]}|\cq)}\leq\frac{d}{2K} = \frac{1}{N}.  
\end{align} 

\textit{Path graph $\mathbb{P}_N$:} To show the upper bound for paths, we need the modified version of \cite[Theorem 6]{SGT23}, to bound the answer size from any server with respect to the answers from servers in its neighbor set, i.e., the servers with which it shares a message and randomness. This is given by the following lemma.

\begin{lemma}\label{lem:single_server_answer_entropy}
    For a server $S\in [N]$, with degree $\delta$ in $G$, let $\cn(S)=\{S_1,\ldots,S_{\delta}\}$ denote its neighbor set. Then, for any $k\in [K]$,
    \begin{align}\label{eq:lem_ineq}
        H(A_S^{[k]}|\cq)\geq \sum_{i=1}^{\delta}\max\left\{0,2L-\sum_{j=i}^\delta H(A_{S_j}^{[k]}|\cq)\right\}.
    \end{align}
\end{lemma}

\begin{Proof} 
In this proof, for every $i\in [\delta]$, let $W_{i}$ and $R_{i}$, respectively, denote the message and randomness stored on servers $S$ and $S_i\in \cn(S)$. Let $\cw^{c}\defeq\cw\setminus \{\cup_{i=1}^{\delta}{W}_i\}$ and $\cR^c\defeq\cR\setminus \{\cup_{i=1}^{\delta}R_i\}$. Conditioning on $\cw^c$ and $\cR^c$, we obtain,
\begin{align}
    H(A_S^{[k]}|\cq)
     =&H(A_S^{[k]}|\cq,\cw^c,\cR^c)-H(A_S^{[k]}|\cw^c,\cR^c,W_{[\delta]},R_{[\delta]},\cq) \label{eq:follows from A_s deterministic}\\
     =&I(A_S^{[k]};W_{[\delta]},R_{[\delta]}|\cw^c,\cR^c,\cq)\label{eq:change k to i}\\
     =&\sum_{i=1}^{\delta} I(A_S^{[i]};W_i,R_i|W_{[i-1]},R_{[i-1]},\cw^c,\cR^c, \cq)\\
    \geq & \sum_{i=1}^{\delta}\left(2L-H(W_i,R_i|A_S^{[i]},W_{[i-1]},R_{[i-1]},\cw^c,\cR^c,\cq)\right)\label{eq:follows by Thm2},
\end{align}
where the second term in \eqref{eq:follows from A_s deterministic} is $0$ because $A_S^{[k]}$ is deterministic given the conditioning variables, \eqref{eq:change k to i} is due to Lemma~\ref{lem:answers_independent_of_index}, and \eqref{eq:follows by Thm2} follows from Theorem \ref{thm:bound_rho}. Next, using \eqref{eq:R_k_deterministic}, we upper bound the second term for each $i$ in the sum of \eqref{eq:follows by Thm2} by $\sum_{j=i}^{\delta}H(A_{S_j}^{[k]}|\cq)$ since,
\begin{align}
    H(&W_i,R_i|A_S^{[i]},W_{[i-1]},R_{[i-1]},\cw^c,\cR^c, \cq)\notag\\
    =& H(W_i, R_i|A_S^{[i]}, W_{[i-1]}, R_{[i-1]},\cw^c, \cR^c,A_{i:\delta}^{[i]},\cq)+I(A^{[i]}_{i:\delta};W_i, R_i|A_S^{[i]}, W_{[i-1]}, R_{[i-1]},\cw^c, \cR^c,\cq) \\
    =& H(W_i|A_S^{[i]},W_{[i-1]},R_{[i-1]},\cw^c,\cR^c, A_{i:\delta}^{[i]}, \cq)+H(R_i|A_S^{[i]},W_i,W_{[i-1]},R_{[i-1]},\cw^c,\cR^c, A_{i:\delta}^{[i]}, \cq) \notag\\
    &+ I(A^{[i]}_{i:\delta};W_i, R_i|A_S^{[i]}, W_{[i-1]}, R_{[i-1]},\cw^c, \cR^c,\cq) \label{eq: first and second terms zero}\\
    \leq & H(A^{[i]}_{i:\delta}|A_S^{[i]}, W_{[i-1]}, R_{[i-1]},\cw^c, \cR^c,\cq)\\
    \leq & \sum_{j=i}^\delta H(A^{[i]}_{S_j}|\cq) =  \sum_{j=i}^\delta H(A^{[k]}_{S_j}|\cq), \label{eq:follows from user privacy}
\end{align}
where we use the shorthand $A^{[i]}_{i:\delta}\defeq\{A^{[i]}_{S_{i}}, \ldots, A^{[i]}_{S_\delta}\}$, the first term in \eqref{eq: first and second terms zero} is zero since $W_i$ is determined from $A_S^{[i]}, A_{i:\delta}^{[i]}$ and the answers of the remaining servers are functions of $W_{[i-1]}, R_{[i-1]}, \cw^c,\cR^c$ and $\cq$, while the second term is zero since  $R_i$ is deterministic given $A_i^{[i]}, W_i, \cw^c, \cR^c$ and $\cq$ by \eqref{eq:R_k_deterministic}. Finally, we obtain \eqref{eq:lem_ineq} from the non-negativity of the mutual information terms in \eqref{eq:change k to i} and the last equality in \eqref{eq:follows from user privacy} follows from user privacy.
\end{Proof}

To show $\mathscr{C}(\mathbb{P}_N)\leq\frac{1}{N}$, we consider the cases of $N$ even and odd separately.

\textit{If $N$ is even:} Let $g$ be a positive integer such that $N=2g$, hence
\begin{align}\label{eq:applying_adjacency_bound}
    \sum_{n=1}^N H(A_n^{[k]})=&\sum_{j=1}^{g} H(A^{[k]}_{2j-1})+H(A^{[k]}_{2j})
    \geq \sum_{j=1}^g 2L, 
\end{align}
where \eqref{eq:applying_adjacency_bound} follows from \eqref{eq:pair_ans_bound} since servers $(2j-1)$ and $2j$ share $W_{2j-1}$ and $R_{2j-1}$ for each $j$. This gives the required bound if $N$ is even. 

\textit{If $N$ is odd:} Let $N=2g+1$, then
\begin{align}
    \sum_{n=1}^N H(A_n^{[k]})=&H(A_1^{[k]})+H(A_2^{[k]})+H(A_3^{[k]})+\sum_{j=2}^{g} H(A^{[k]}_{2j})+H(A^{[k]}_{2j+1})\\
    \geq& H(A_1^{[k]})+H(A_2^{[k]})+H(A_3^{[k]})+(N-3)L, \label{eq:sum_path answers_odd}
\end{align}
where \eqref{eq:sum_path answers_odd} follows from \eqref{eq:pair_ans_bound} and since $2(g-1)=N-3$. It remains to show that $H(A_1^{[k]})+H(A_2^{[k]})+H(A_3^{[k]})\geq 3L$. If $H(A_3^{[k]})\geq L$, since $H(A_1^{[k]})+H(A_2^{[k]})\geq 2L$, we are done. Otherwise, Lemma \ref{lem:single_server_answer_entropy} applied to $S=2$, yields
\begin{align}
    H(A_2^{[k]})\geq& \max\left\{0,2L-H(A_1^{[k]})-H(A_3^{[k]})\right\}+ 2L-H(A_3^{[k]})\label{eq:odd_path_reduction}, 
\end{align}
since $\cn(S)=\{1,3\}$. Then, \eqref{eq:odd_path_reduction} reduces to
\begin{align}\label{eq:bound_A2}
    H(A_2^{[k]})\geq 
    \begin{cases}
    2L-H(A_3^{[k]}), & \text{ if } H(A_1^{[k]})+H(A_3^{[k]})\geq 2L,\\
    4L-H(A_1^{[k]})-2H(A_3^{[k]}), & \text{ otherwise.}
    \end{cases}
\end{align}
Rearranging the terms in \eqref{eq:bound_A2} yields $H(A^{[k]}_1) +H(A_2^{k]}) +H(A_3^{[k]})\geq 3L$ in both cases, which by substitution in \eqref{eq:sum_path answers_odd} gives the required bound for $N$ odd. 

\subsection{SPIR Schemes from PIR Schemes}\label{sec:spir from pir gr} 
In this section, we propose a strategy to convert a class of PIR schemes on $G$ into its respective SPIR schemes. The resulting schemes allow a flexible transition from a PIR scheme on $G$ into an SPIR scheme, upon requirement of database privacy and availability of graph-replicated common randomness. The scheme construction proceeds in two stages, and consequently requires the underlying PIR scheme to satisfy SRP. We show that, the SPIR rate achieved through this construction is exactly the half of the rate of the respective PIR scheme. 

Our construction focuses on the class of PIR schemes, where the servers' answers are downloaded in the form of $l$-sums, i.e., sums of symbols of distinct messages stored in them, with $l\geq1$. Let $\theta$ denote the desired message index, and let $T$ denote a PIR scheme on $G$ which satisfies SRP. Let $L'$ and $D'$ denote the number of symbols per message, and the number of downloaded symbols, respectively, required for $T$. Thus, $R_{PIR}=\frac{L'}{D'}$ denotes the rate of the PIR scheme $T$. To build the SPIR scheme on top of $T$, we let $L=L'$ and $\rho=1$. 

We start with motivating examples on path, cyclic and star graphs.

\begin{example}[Path graph]\label{ex:scheme2_path}
Table \ref{tab:pir_p3} shows the answer table for the PIR scheme on $\mathbb{P}_3$ \cite{our_journal2025} with $W_1=(a_1,a_2)$ and $W_2=(b_1,b_2)$, representing the messages after the user has privately permuted the symbols. Since a single symbol of desired message is retrieved from the servers where it is replicated, the scheme satisfies SRP. 

To build the SPIR scheme on top of this, assume that $L=2$, and the randomness symbols are $R_1 = (s_{1,1}, s_{1,2})$ and $R_2 = (s_{2,1}, s_{2,2})$. The answer table for our SPIR scheme in shown in Table~\ref{tab:scheme2_path}, which can be viewed as a two-step scheme. In the first step, the user and servers conduct the PIR scheme in Table \ref{tab:pir_p3} on $\{R_1,R_2\}$ to privately retrieve the desired $R_k$. In the second step, the user and servers conduct the PIR scheme on $W_1,W_2$, added to $R_1,R_2$ to retrieve the desired $W_k+R_k$. The rate is $\frac{1}{3}$, also the capacity.
\end{example}

\begin{table}[h]
    \centering
    \begin{tabular}{|c|c|c|c|}
    \hline
    & server 1 & server 2 & server 3\\
    \hline
    {$k = 1$} 
    & $a_1$ & $a_2+b_2$ & $b_2$\\
    \hline
    {$k = 2$} 
    & $a_1$ & $a_1+b_1$ & $b_2$\\
    \hline
    \end{tabular}
    \caption{PIR answer table for $\mathbb{P}_3$ \cite{our_journal2025}.}
    \label{tab:pir_p3}
\end{table}

\begin{table}[h]
    \centering
    \begin{tabular}{|c|c|c|c|}
    \hline
    & server 1 & server 2 & server 3\\
    \hline
    \multirow{2}{*}{$k = 1$} 
    & $s_{1,1}$ & $s_{1,2}+s_{2,2}$ & $s_{2,2}$\\
    & $a_1+s_{1,2}$ & $a_2+b_2+s_{1,1}+s_{2,1}$ & $b_2+s_{2,1}$\\
    \hline
    \multirow{2}{*}{$k = 2$} 
    & $s_{1,1}$ & $s_{1,1}+s_{2,1}$ & $s_{2,2}$\\
    & $a_1+s_{1,2}$ & $a_1+b_1+s_{1,2}+s_{2,2}$ & $b_2+s_{2,1}$\\
    \hline
    \end{tabular}
    \caption{SPIR answer table for $\mathbb{P}_3$.}
    \label{tab:scheme2_path}
\end{table}

\begin{example}[Cyclic graph] Table~\ref{tab:pir_c3} shows the PIR scheme for the cyclic graph with $W_1= (a_1,\ldots,a_6)$, $W_2 = (b_1,\ldots,b_6)$ and $W_3 = (c_1,\ldots,c_6)$ denoting the permuted message symbols \cite{BU19}.\footnote{In this example, we presented the capacity-achieving scheme for cyclic graphs in a reduced form. With respect to the original scheme, the number of downloads and the number of desired symbols are reduced by a factor of $\frac{1}{N-1}$ without affecting the PIR requirements and achievable rate.} Note that the PIR rate is $\frac{6}{3\times 4}=\frac{1}{2}$. Let $R_1 = (s_{1,1},\ldots,s_{1,6})$, $R_2 = (s_{2,1},\ldots,s_{2,6})$ and $R_3 = (s_{3,1}, \ldots, s_{3,6})$. As shown in Table~\ref{tab:scheme2_cycle}, the user retrieves the required randomness $R_1$ through the PIR scheme on $\{R_1,R_2,R_3\}$. This is followed by the user retrieving $W_1+R_1$ through the same PIR scheme, this time applied to the sums of message and randomness symbols. The SPIR rate is $\frac{6}{3\times 8} = \frac{1}{4}$, less than the rate in Example~\ref{ex:cyclic graph general scheme}.
\end{example}

\begin{table}[h]
    \centering
    \begin{tabular}{|c|c|c|}
    \hline
    server 1 & server 2 & server 3\\
    \hline
    $a_1, c_1$ & $a_4, b_1$ & $b_2, c_2$\\
    $a_2+c_2$ & $a_5+b_2$ & $b_3+c_1$\\
    $a_3 + c_3$ & $a_6+b_3$ & $b_1+c_3$\\
    \hline
    \end{tabular}
    \caption{PIR answer table for $\mathbb{C}_3$ \cite{BU19} when $k = 1$.}
    \label{tab:pir_c3}
\end{table}

\begin{table}[h!]
    \centering
    \begin{tabular}{|c|c|c|}
    \hline
    server 1 & server 2 & server 3\\
    \hline
    $s_{1,1},s_{3,1}, s_{1,2}+s_{3,2}, s_{1,3}+s_{3,3}$ & $s_{1,4},s_{2,1}, s_{1,5}+s_{2,2}, s_{1,6}+s_{2,3}$ & $s_{2,2},s_{3,2}, s_{2,3}+s_{3,1}, s_{2,1}+s_{3,3}$\\
    $a_1+s_{1,4}, c_1+s_{3,4}$ & $a_4+s_{1,1}, b_1+s_{2,4}$ & $b_2+s_{2,5}, c_2+s_{3,5}$\\
    $a_2+c_2+s_{1,5}+s_{3,5}$ & $a_5+b_2+s_{1,2}+s_{2,5}$ & $b_3+c_1+s_{2,6}+s_{3,4}$\\
    $a_3 + c_3 +s_{1,6} +s_{3,6}$ & $a_6+b_3+s_{1,3}+s_{2,6}$ & $b_1+c_3+s_{2,4}+s_{3,6}$\\
    \hline
    \end{tabular}
    \caption{SPIR answer table for $\mathbb{C}_3$ when $k = 1$.}
    \label{tab:scheme2_cycle}
\end{table}

\begin{example}[Star graph]
     We let $L=2$ symbols per message and $R_1=(s_{1,1},s_{1,2})$, $R_2 = (s_{2,1},s_{2,2})$ and $R_3 = (s_{3,1},s_{3,2})$. Then, we build our SPIR scheme on a simple PIR scheme \cite{YJ23} (see Table~\ref{tab:pir_star}). Note that the PIR rate is $\frac{2}{4}=\frac{1}{2}$. The corresponding SPIR answer table is given in Table \ref{tab:spir_scheme2_s4}, and the rate achieved here is $\frac{1}{4}$ with $\rho=1$, equal to those in Example~\ref{ex:star graph general scheme}.
\end{example}

\begin{table}[ht]
    \centering
    \begin{tabular}{|c|c|c|c|c|}
    \hline
    & server 1 & server 2 & server 3 & server 4\\
    \hline
    {$k = 1$} 
    & $a_1$ & $b_1$ & $c_1$ & $a_2+b_1+c_1$\\
    \hline
    {$k = 2$} 
    & $a_1$ & $b_1$ & $c_1$ & $a_1+b_2+c_1$\\
    \hline
    {$k = 3$} 
    & $a_1$ & $b_1$ & $c_1$ & $a_1+b_1+c_2$\\
    \hline
    \end{tabular}
    \caption{PIR answer table for $\mathbb{S}_4$ \cite{YJ23}.}
    \label{tab:pir_star}
\end{table}

\begin{table}[ht]
    \centering
    \begin{tabular}{|c|c|c|c|c|}
    \hline
    & server 1 & server 2 & server 3 & server 4\\
    \hline
    \multirow{2}{*}{$k = 1$} 
    & $s_{1,1}$ & $s_{2,1}$ & $s_{3,1}$ & $s_{1,2}+s_{2,1}+s_{3,1}$\\
    & $a_1+s_{1,2}$ & $b_1+s_{2,2}$ & $c_1+s_{3,2}$ & $a_2+b_1+c_1+s_{1,1}+s_{2,2}+s_{3,2}$\\
    \hline
    \multirow{2}{*}{$k = 2$} 
    & $s_{1,1}$ & $s_{2,1}$ & $s_{3,1}$ & $s_{1,1}+s_{2,2}+s_{3,1}$\\
    & $a_1+s_{1,2}$ & $b_1+s_{2,2}$ & $c_1+s_{3,2}$ & $a_1+b_2+c_1+s_{1,2}+s_{2,1}+s_{3,2}$\\
    \hline
    \multirow{2}{*}{$k = 3$} 
    & $s_{1,1}$ & $s_{2,1}$ & $s_{3,1}$ & $s_{1,1}+s_{2,1}+s_{3,2}$\\
    & $a_1+s_{1,2}$ & $b_1+s_{2,2}$ & $c_1+s_{3,2}$ & $a_1+b_1+c_2+s_{1,2}+s_{2,2}+s_{3,1}$\\
    \hline
    \end{tabular}
    \caption{SPIR answer table for $\mathbb{S}_4$.}
    \label{tab:spir_scheme2_s4}
\end{table}

Because $T$ admits SRP, $L$ is even. It is clear from the examples that the main idea is to apply $T$ in two steps: first, to retrieve $R_k$ and second, to retrieve $W_k+R_k$, as described next. 

\textit{Index Permutation:} As specified by $T$, the user privately chooses independent permutations on the symbol indices of $W_\ell$, $\ell \in [K]$. In addition, the user applies the same permutations on the respective randomness symbol indices of $R_{\ell}$, $\ell\in [K]$, with the permuted symbols denoted as $R_\ell = (s_{\ell,1},s_{\ell,2},\ldots,s_{\ell,L})$. 

\textit{Query Generation:} The query structure of SPIR follows that of $T$, with the desired message index $k$. Consider the query sent to server $n\in [N]$. Let, for each $\ell\in \mathcal{F}_n$, $P_{\ell,n}\subsetneq [L]$ denote the permuted symbol indices of $W_\ell$ that are queried from server $n$. By SRP of $T$ and user privacy, it holds that, $|P_{\ell,n}|=\frac{L}{2}$ for all $n$. We define a bijection $\psi:[L]\to [L]$ with
\begin{align}
    \psi(m) =
    \begin{cases}
        m+\frac{L}{2}, & m\in [1:\frac{L}{2}],\\
        m-\frac{L}{2}, & m\in [\frac{L}{2}+1:L].
    \end{cases}
\end{align}
Independent of $\theta$, the mapping shifts the queried indices by $\frac{L}{2}$. Along with the PIR queries, the user sends the  mapping $\psi$ to the servers.  

\textit{Answer Generation:} The servers respond with two sets of answers. First, the servers respond by applying $T$ to the randomness $\mathcal{R}$ that they store.  In particular, the answers of server $n$ are composed of the randomness symbols $\{s_{\ell,\mu}, \mu\in P_{\ell,n}, \ell\in \mathcal{F}_n\}$. Second, corresponding to each query of $T$, the servers first apply $T$ to the symbols of $\cw$, creating the answer table of the PIR scheme $T$. Each answer is added to the corresponding answer of $T$ applied to symbols of $\cR$, mapped under $\psi$. That is, to every message symbol of $W_\ell$, with index $\mu\in P_{\ell,n}$ in the PIR answer table, server $n$ adds the randomness symbol $s_{\ell, \psi(\mu)}$ of $R_{\ell}$, for each $\ell\in \mathcal{F}_n$.

Next, we argue that our construction satisfies the requirements, \eqref{eq:user_privacy}, \eqref{eq:reliability} and \eqref{eq:database_privacy1}.

\textit{Reliability:} From the first set of answers, the user recovers the $L$ symbols of $R_k$, by the reliability of scheme $T$. From the second set of answers, the user recovers $L$ symbols of $W_k$ added to $L$ symbols of $R_k$. Using these and $\psi$, the user decodes $W_k$, by canceling out the respective symbols of $R_k$ from the first set. Therefore, our construction guarantees reliability.

\textit{User Privacy:} This follows from the privacy of the individual PIR schemes on $\cR$ and $\cw + \cR$, the injectivity of $\psi$ and the private permutations applied to the message and randomness symbol indices. Our scheme ensures that, from every server, the user downloads equal number of all its stored message and randomness symbols as answers.

\textit{Database Privacy:} In the scheme, the user downloads randomness symbols in the first set of answers, which are either of $R_k$, or of $R_{\ell}$, $\ell\in [K]\setminus \{k\}$, where the latter serve as interference symbols to retrieve $R_k$. Also, the symbols of $R_{\ell}$ downloaded in the first set are distinct from those added to $W_\ell$, $\ell\in [K]\setminus \{k\}$, in the second set, by definition of $\psi$. Thus, $R_\ell$ for $\ell\in [K]\setminus \{k\}$ are unavailable to the user, because of which $W_{\overline{k}}$ remains private.

We see that, the number of downloaded symbols in each answer set is $D'=\frac{L}{R_{PIR}(G)}$. This entails twice the download cost of PIR scheme $T$. With $L$ symbols of $W_k$ retrieved, the resulting SPIR rate is $R(G)=\frac{R_{PIR}(G)}{2}$.

\section{SPIR with Fully-Replicated Randomness}\label{sec:fully-replicated simple}
In this section, we consider the variant of SPIR, where even though the messages are graph-replicated, $\cR$ is fully-replicated at all servers. Towards this, we derive bounds on $\mathscr{C}_{FR}(G)$ and $\rho^*_{total}(G)$. The lower bounds on $\mathscr{C}_{FR}(G)$ are derived from PIR-induced schemes on specific graphs. Such schemes enable the user and the servers to transition into the SPIR scheme, provided that $\cR$ is fully-replicated. 

\subsection{SPIR from PIR Schemes}
We prove Theorem~\ref{thm:spir scheme fr} by presenting the algorithm that converts a PIR scheme into an SPIR scheme. Then, we focus on a simple construction to extract an SPIR scheme, given a PIR scheme on the star graph $\mathbb{S}_N$. Unlike the previous approach, this construction does not require SRP of the PIR scheme, and is feasible due to the specific replication pattern of star graphs; namely, a specific server stores the entire message set $\cw$, and each remaining server stores a single distinct message of $\cw$. For $N=3$, this SPIR scheme achieves the capacity $\mathscr{C}_{FR}$ for $\mathbb{P}_3$, proving the achievability of Theorem~\ref{thm:capacity of P_3 fr}. 

We motivate the first construction with two examples.

\begin{example}[Path graph]\label{ex:path}
Consider the SPIR system for $\mathbb{P}_3$ as shown in Fig.~\ref{fig:sysmod_fr}(a). Suppose, $L=4$ and $H(\cR)=5$. Let the message and randomness symbols, after private and independent permutations by the user be denoted by $W_1=(a_1,\ldots,a_4)$, $W_2=(b_1,\ldots,b_4)$ and $\cR=(s_1,\ldots,s_5)$. The answers of the SPIR scheme, built on the PIR scheme in Table~\ref{tab:pir_p3} are shown in Table~\ref{tab:spir_p3_fully_rep}. First, the user downloads a unique randomness symbol from each server. Next, the user performs two repetitions of the PIR scheme on $\mathbb{P}_3$, with the modification that each downloaded answer is mixed with randomness symbols from $\cR$. For instance, if $k=1$, we add a new randomness symbol $s_3,s_5$ to $b_2$ and $b_4$, respectively. Similarly, we add $s_1$ (downloaded from server 2) to $a_1$, and $s_2$ (downloaded from server 1) to $a_2$, and $s_4$ (downloaded from server 3) to both $a_3$ $a_4$. This gives the rate $\frac{4}{9}=\frac{2}{3+\frac{3}{2}}$ and $\rho_{total}(\mathbb{P}_3)=\frac{5}{4}=\frac{9}{4}-1$.
\end{example}

\begin{table}[ht]
    \centering
    \begin{tabular}{|c|c|c|c|}
    \hline
    $k=1$ & server 1 & server 2 & server 3\\
    \hline
    &$s_2$ & $s_1$ & $s_4$\\
    rep. 1&$a_1+s_1$ & $a_2+b_2+s_2+s_3$ & $b_2+s_3$\\
    rep. 2&$a_3+s_4$ & $a_4+b_4+s_4+s_5$ & $b_4+s_5$ \\
    \hline
    \hline
    $k=2$ & server 1 & server 2& server 3\\
    \hline
    &$s_5$ & $s_3$ & $s_2$\\
    rep. 1&$a_1+s_1$ & $a_1+b_1+s_1+s_2$ & $b_2+s_3$\\
    rep. 2&$a_3+s_4$ & $a_3+b_3+s_4+s_5$ & $b_4+s_5$ \\
    \hline
    \end{tabular}
    \caption{SPIR answer table for $\mathbb{P}_3$.}
    \label{tab:spir_p3_fully_rep}
\end{table}

\begin{example}[Cyclic graph]
For the SPIR scheme on $\mathbb{C}_N$ with $N=3$, we build upon the scheme in \cite{BU19} (also shown in Table~\ref{tab:pir_c3}) to form the answer table when $W_1$ is desired, as shown in Table~\ref{tab:spir_c3_fully_rep}. Corresponding to each symbol of $W_2,W_3$, we assign a new distinct randomness symbol to be added to the answer. For each symbol of $W_1$ downloaded from server 1 (or, server 2), a randomness symbol downloaded from server 2 (or, server 1) and server 3 is assigned. The resulting rate is $\frac{4}{11}=\frac{2}{3+1+\frac{3}{2}}$ and $\rho_{total}(\mathbb{C}_3)=\frac{7}{4}=\frac{11}{4}-1$.
\end{example}

\begin{table*}[ht]
    \centering
    \begin{tabular}{|c|c|c|c|}
    \hline
    & server 1 & server 2 & server 3\\
    \hline
    & $s_7,s_9,s_{11}$ & $s_1,s_3,s_5$ & $s_{13},s_{15},s_{17}$\\
    \hline
    \parbox[t]{2mm}{\multirow{3}{*}{\rotatebox[origin=c]{90}{rep. 1}}} &$a_1+s_1,c_1+s_2$ & $a_4+s_7, b_1+s_8$ & $b_2+s_{10}, c_2+s_4$\\
    & $a_2+c_2+s_3+s_4$ & $a_5+b_2+s_9+s_{10}$ & $b_1+c_3+s_8+s_6$\\
    & $a_3+c_3+s_5+s_6$ & $a_6+b_3+s_{11}+s_{12}$ & $b_3+c_1+s_{12}+s_2$\\
    \hline
    \parbox[t]{2mm}{\multirow{3}{*}{\rotatebox[origin=c]{90}{rep. 2}}}  &$a_7+s_{13},c_7+s_{14}$ & $a_{10}+s_{13}, b_7+s_{19}$ & $b_8+s_{20}, c_8+s_{16}$\\
    & $a_8+c_8+s_{15}+s_{16}$ & $a_{11}+b_8+s_{15}+s_{20}$ & $b_7+c_9+s_{19}+s_{18}$\\
    & $a_9+c_9+s_{17}+s_{18}$ & $a_{12}+b_9+s_{17}+s_{21}$ & $b_9+c_7+s_{14}+s_{21}$\\
    \hline
    \end{tabular}
    \caption{SPIR answer table for $\mathbb{C}_3$ when $k=1$.}
    \label{tab:spir_c3_fully_rep}
\end{table*}

\subsubsection{Algorithm Description}\label{subsec:achievable_algo}
It is clear from the examples that our scheme exhibits two key properties: 1) The user downloads an equal number $y$ of uncoded randomness symbols from every server, and 2) if the randomness symbols are removed from the answers, our scheme reduces to $x$ repetitions of the starting PIR scheme.

Let $W_k$ be the desired message, replicated on servers $i, j\in [N]$. To determine $x$ and $y$, for a given PIR scheme, observe that, in each repetition $x$, the user retrieves $xL'/2$ symbols of $W_k$, masked by a symbol of $\cR$, from each server $i$ and $j$. For decoding the $xL'/2$ symbols of $W_k$ retrieved from each of server $i$ and $j$, the user downloads the corresponding $xL'/2$ randomness symbols from the remaining servers, i.e., $[N]\setminus \{i\}$ and $[N]\setminus \{j\}$, respectively. By property 1, this implies that, $x$ and $y$ are the least positive integers satisfying,
\begin{align}\label{eq:relation of x and y}
    x\frac{L'}{2}=(N-1)y.
\end{align}
Clearly, \eqref{eq:relation of x and y} is equal to $\lcm(L'/2,N-1)$ by the definition of least common multiple. Now, we describe our algorithm steps.

\textit{Index Permutation:} For the SPIR scheme, let $L=xL'$ and $H(\cR)=Ny+\left(\frac{K-1}{2}\right)L$. The user applies private and independent permutations to the symbols of the $K$ messages, with $W_\ell=\left(w_\ell(1), \ldots, w_\ell(L)\right)$ and $\cR = (s_1,\ldots,s_{Ny+L(K-1)/2})$.

\textit{Query Generation:} The user follows the query structure of the starting PIR scheme. Queries sent to the servers correspond to $x$ repetitions of the scheme, where in each repetition, the user queries for $L'$ new symbols of $W_{k}$. Corresponding to each queried symbol, the user sends the following common randomness assignment to each server. 

\textit{Common Randomness Assignment:} The user assigns a unique randomness symbol to each queried message symbol $W_{\ell}\in W_{\overline{k}}$ across all repetitions. Note that, because of SRP, to maintain user privacy, we query $L'/2$ symbols of $W_\ell$ from every server in the PIR scheme. For reliability of the PIR scheme through interference cancellation, we require the same\footnote{If a non-desired message symbol is queried only for privacy preservation, and does not participate in interference cancellation, (e.g., scheme for $\mathbb{K}_N$ in \cite{our_journal2025}) we assume without loss of generality that these are identical for both servers.} $L'/2$ symbols to be queried from both servers where $W_\ell$ is replicated. This assigns the $(K-1)L/2$ symbols of $\cR$. To the first $y=\frac{L}{2(N-1)}$ symbols of $W_k$ queried from server $i$ and $j$, the user assigns distinct randomness symbols $s_{j_1},\ldots,s_{j_y}$ and $s_{i_1},\ldots,s_{i_y}$, respectively. To each of the remaining $L/2-y=(N-2)y$ queried symbols of $W_k$ from each of server $i$ and $j$, the user assigns the same randomness symbol from the remaining set $\{s_{n_1},\ldots,s_{n_y}, n\in [N]\setminus\{i,j\}\}$ of $\cR$.  

\textit{Answer Formation:} The user downloads the randomness symbols $s_{n_1},\ldots,s_{n_y}$ from server $n\in [N]$, followed by $x$ repetitions of PIR answers. To each answer, the servers add randomness symbols, with each message symbol in the answer added to the assigned randomness symbol. That is, a $t$-sum in the answer is padded with $t$ symbols from $\cR$.

This completes the algorithm, with $R_{FR}$ and $\rho_{total}$ derived by direct computation. Next, we show that our algorithm outputs a feasible SPIR scheme.

\textit{Reliability:} By reliability of the PIR scheme, in repetition $u\in [x]$, the user retrieves $w_k((u-1)L'+1),\ldots,w_k(u L')$ added to an already downloaded symbol of $\cR$. Using this, they recover $L=xL'$ symbols of $W_k$. Specifically, $s_{j_1},\ldots,s_{j_y}$ and $s_{i_1},\ldots,s_{i_y}$ cancel out the randomness symbols to retrieve the first $y$ symbols of $W_k$ from server $i$ and $j$, respectively. Further, $s_{n_1},\ldots,s_{n_y}$ downloaded from servers $n\in[N]\setminus \{i,j\}$ are used as common side information to retrieve the remaining $L-2y$ symbols of $W_k$ from both servers $i$ and $j$. 

\textit{User Privacy:} From the perspective of each server, the queried message symbols are independent of the desired index $k$, due to the PIR scheme. Further, the randomness symbols directly downloaded by the user, and those added to the PIR answers, appear to be uniformly chosen from $\cR$, irrespective of $k$, since the private permutation on $\cR$ hides the common randomness assignment.

\textit{Database Privacy:} No information on $W_{\overline{k}}$ is revealed to the user, since every non-desired message symbol is protected by a unique symbol of $\cR$, which is distinct from those downloaded by the user in the scheme. 

\subsubsection{SPIR Schemes on Star Graphs}\label{subsec:star scheme}
A star graph $\mathbb{S}_N$ with $N$ vertices and $N-1$ edges is described by the storage pattern where,  for every $n\in [N-1]$, $W_n$ stored at server $n$ and the full message set $\cw=\{W_1,\ldots,W_{N-1}\}$ is stored at server $N$. For $N=4$, adopting the algorithm in Section~\ref{subsec:achievable_algo} as shown in Table~\ref{tab:suboptimal scheme star} for $k=1$ yields the rate $R_{FR}(\mathbb{S}_4)=\frac{3}{8}$ and $\rho_{total}(G) = \frac{5}{3}$. This is derived from a PIR scheme for $\mathbb{S}_N$ with $L'=2$ and $D'=N$. In particular, for $k\in [N-1]$, the user, after permuting the message and randomness symbols, queries for a single message symbol of $W_n$ from servers $n\in [N-1]$, and the sum of a new symbol of $W_{k}$ added to sum of the queried interference symbols $\cw\setminus \{W_k\}$. For this PIR scheme, $x=N-1$, $y=1$ and $R_{FR}(\mathbb{S}_{N})=\frac{2}{N+\frac{N}{N-1}}$ with $\rho_{total}(\mathbb{S}_N) = \frac{N-1}{2}+\frac{N}{2(N-1)}$. In the following, we construct a scheme with superior SPIR rate.

The PIR capacity $\mathscr{C}_{PIR}(\mathbb{S}_N)$ was approximately characterized in \cite{SGT23} to be $\Theta(\frac{1}{\sqrt{N}})$ for large $N$. The combinatorial design-based scheme proposed in \cite{SGT23} was applicable only when $K=N-1$ is a perfect square. The resulting lower bound was sharpened through a general scheme, proposed in \cite{YJ23} for $\mathbb{S}_N$, which does not satisfy SRP in general. This is why a multi-round SPIR scheme as in Section~\ref{subsec:achievable_algo} is not feasible. However, all servers except server $N$ store a single message, enabling the construction of a simple SPIR scheme from any PIR scheme on $\mathbb{S}_N$. For instance, Table~\ref{tab:better scheme star} shows the SPIR answers for $N=4$, where the underlying PIR scheme does not satisfy SRP. This scheme achieves a higher rate of $R_{FR}(\mathbb{S}_4)=\frac{3}{7}$ with $\rho_{total}(\mathbb{S}_N)=\frac{4}{3}$. 

\begin{table}[t]
    \centering
    \begin{tabular}{|c|c|c|c|c|}
    \hline
    & server 1 & server 2 & server 3 & server 4\\
    \hline
    & $s_4$ & $s_5$ & $s_8$ & $s_1$\\
    \hline
    rep. 1 &$a_1+s_1$ & $b_1+s_2$ & $c_1+s_3$ & $a_2+b_1+c_1+s_4+s_2+s_3$\\
    \hline
    rep. 2 &$a_3+s_5$ & $b_3+s_6$ & $c_3+s_7$ & $a_4+b_3+c_3+s_5+s_6+s_7$\\
    \hline
    rep. 3 &$a_5+s_8$ & $b_5+s_9$ & $c_5+s_{10}$ & $a_6+b_5+c_5+s_8+s_9+s_{10}$\\
    \hline
    \end{tabular}
    \caption{SPIR answer table on $\mathbb{S}_4$ ($k=1$) using scheme in Section~\ref{subsec:achievable_algo};  rate $\frac{3}{8}$.}
    \label{tab:suboptimal scheme star}
\end{table}
    
\begin{table}[t]
    \centering
    \begin{tabular}{|c|c|c|c|}
    \hline
    server 1 & server 2 & server 3 & server 4\\
    \hline
      & & & $s_1$\\
    \hline
    $a_1+s_1$ & $b_1+s_2$ & $c_1+s_3$ & $a_2+b_1+s_2, a_3+c_1+s_3, b_2+c_2+s_4$\\
    \hline
    \end{tabular}
    \caption{SPIR answer table for $\mathbb{S}_4$ ($k=1$) using scheme in Sectiom~\ref{subsec:star scheme}; rate $\frac{3}{7}$.}
    \label{tab:better scheme star}
\end{table}

Let $W_k$ be the desired message. The scheme in \cite{YJ23} is parameterized by the integer $t\in [N-1]$, where for a given $t$, the user queries the following. From server $N$, the user queries $\binom{K}{t}=\binom{N-1}{t}$ $t$-sums of $\cw$, each corresponding to a $t$-sized subset of $[N-1]$. Each $t$-sum comprises new message symbols of distinct messages, i.e., $\binom{N-2}{t-1}$ symbols of every message $W_\ell, \ell\in [N-1]$. Among these, each non-desired message $W_{\ell}$, $\ell\in [K]\setminus \{k\}$ is added to $W_k$ in $\binom{N-3}{t-2}$ $t$-sums. From servers $n\in [N-1]$, the user queries for $\binom{N-3}{t-2}$ new symbols of $W_k$,  if $n=k$; and $\binom{N-3}{t-2}$ symbols of $W_n$, that are summed with $W_k$ in the answers of server $N$, if $n\neq k$. This amounts to, 
\begin{align}
    L'&=\binom{N-2}{t-1}+\binom{N-3}{t-2} \\
    D'&=(N-1)\binom{N-3}{t-2}+\binom{N-1}{t},    
\end{align}
resulting in a series of achievable rates, parametrized by $t$. Taking the maximum over $t$ yields the rate, 
\begin{align}
    R_{PIR}(\mathbb{S}_N)=\max_{t\in [1:N-1]}\frac{\binom{N-2}{t-1}+\binom{N-3}{t-2}}{(N-1)\binom{N-3}{t-2}+\binom{N-1}{t}}.
\end{align}

The SPIR scheme built on top of this, requires $L=L'$ and $H(\cR)=\binom{N-3}{t-2}+(t-1)\binom{N-1}{t}$, where $t\in [2:N-1]$ is an integer. The PIR answers are modified to yield SPIR answers as follows. From servers $n\in [N-1]$, for each distinct queried message symbol, we assign a distinct symbol of $\cR$. These servers mask their PIR answers (individual message symbols) with the corresponding randomness symbols. From server $N$, the user requests for two kinds of downloads. First, the user downloads the  $\binom{N-3}{t-2}$ randomness symbols added to the symbols of $W_k$ in the answers of server $k$. Second, the user downloads the usual $t$-sums in the PIR answers, each added to a unique $(t-1)$-sum of randomness symbols. In particular, each $t$-sum belongs to one of the following types: 1) The $t$-sum involves a symbol of $W_k$ added to $(t-1)$ non-desired message symbols; there are $\binom{N-2}{t-1}$ such sums. Since these participate in decoding $W_k$, we add the same $t-1$ randomness symbols that mask the respective non-desired message symbols from servers $n\in [N-1]\setminus \{k\}$. This accounts for $(N-2)\binom{N-3}{t-2}$ symbols of $\cR$. 2) The $t$-sum does not involve a symbol of $W_k$, i.e., it comprises $t$ non-desired message symbols; there are $\binom{N-2}{t}$ such sums. Since these $t$-sums do not participate in decoding $W_k$, they must be protected from the user for database privacy. Although adding a single randomness symbol to each such sum would suffice to protect database privacy, the non-uniformity in answer formation breaches user privacy. Consequently, the scheme demands adding a unique $(t-1)$-sum of unused randomness symbols to each such sum. This accounts for the remaining $(t-1)\binom{N-2}{t}$ symbols of $\cR$. 

This completes the SPIR scheme. For a given $t$, the SPIR rate is given by,
\begin{align}
    R_{FR}(\mathbb{S}_N) & = \max_{t\in [2:N-1]}\frac{L'}{D'+\binom{N-3}{t-2}} \notag \\
    & =  \max_{t\in [2:N-1]}\frac{\binom{N-2}{t-1}+\binom{N-3}{t-2}}{N\binom{N-3}{t-2}+\binom{N-1}{t}} \\
    & \geq \frac{2(N-1)}{2N+(N-1)(N-2)} \label{eq:substitute t=2}\\
    & > \frac{2}{N+\frac{N}{N-1}}, \label{eq:since N exceeds 2}
\end{align}
where \eqref{eq:substitute t=2} follows by substituting $t=2$, and \eqref{eq:since N exceeds 2} holds since $N\geq 3$. The $\rho_{total}(\mathbb{S}_N)$ is $\frac{H(\cR)}{L}$ at the $t$ which achieves $R_{FR}(\mathbb{S}_N)$.

For $N=3$, the star graph is identical to the path graph $\mathbb{P}_3$, for which the aforementioned scheme achieves the capacity $\mathscr{C}_{FR}(\mathbb{P}_3)=\frac{1}{2}$, with $\rho^*_{total}(\mathbb{P}_3)=1$. We show this by proving a matching converse in Section~\ref{sec:proof_upperbound path and cycles}.

\subsection{Capacity Upper Bounds}\label{sec:proof_upperbound path and cycles}
In the following, we derive upper bounds on the SPIR capacity of path and cyclic graphs under the fully-replicated common randomness setting. We also establish a lower bound on $\rho_{total}$ for these graphs. First, we state the lemmas which will be used in the proofs. 

The first lemma is a consequence of user privacy \eqref{eq:user_privacy} and graph-replicated database. 

\begin{lemma}\label{lem: conseq of user priv}
    For any $\mathcal{J}\subseteq [K]$, and any server $n\in [N]$, for all $k, k'\in [K]$, we have
    \begin{align}
        H(A_n^{[k]}|W_\mathcal{J},\cR,Q_n^{[k]}) &= H(A_n^{[k']}|W_\mathcal{J},\cR,Q_n^{[k']}),\\
            H(A_n^{[k]}|W_\mathcal{J},Q_n^{[k]}) &= H(A_n^{[k']}|W_\mathcal{J},Q_n^{[k']}).
    \end{align}
\end{lemma}

The next lemma generalizes \cite[Lemma 2]{c_spir} and is a variation of Lemma~\ref{lem:answer_indep_randomness_given_query} of the graph-replicated randomness setting.

\begin{lemma}\label{lem:answer n depends only on query n}
    For any $\mathcal{J}\subseteq [K]$ and $k\in [K]$, it holds that
    \begin{align}
        H(A_n^{[k]}|W_{\mathcal{J}},\cR, Q_n^{[k]})=H(A_n^{[k]}|W_{\mathcal{J}},\cR, Q_n^{[k]},\cq).
    \end{align}
\end{lemma}

The following lemma is an extension of \cite[Lemma 3]{SGT23} for database privacy \eqref{eq:database_privacy2}, and also a variation of Lemma~\ref{lem:sum_entropy of answers}.

\begin{lemma}\label{sum of answer entropies}
For $k\in [K]$, let $W_k$ be replicated on servers servers $i$ and $j$. Then, 
    \begin{align}
        H(A_i^{[k]}|W_{\overline{k}},\cR,\cq) + H(A_j^{[k]}|W_{\overline{k}},\cR,\cq)\geq L.
    \end{align}
\end{lemma}

\begin{Proof}
From reliability and independence of $\cw$ and $\cq$,  it follows that
\begin{align}
    L&=H(W_k|\cq)-H(W_k|A^{[k]}_{1:N},\cq)\\
     &\leq H(W_k|W_{\overline{k}},\cR,\cq) - H(W_k|W_{\overline{k}},A^{[k]}_{1:N},\cR,\cq)\label{eq:independence of messages fr}\\
     &= H(W_k|W_{\overline{k}},\cR,\cq) - H(W_k|W_{\overline{k}},A^{[k]}_i,A^{[k]}_j,\cR,\cq)\label{eq: all except Ai and Aj are deterministic}\\
     &= I(A_i^{[k]},A_j^{[k]};W_k|W_{\overline{k}},\cR,\cq)\\
     &=H(A_i^{[k]},A_j^{[k]}|W_{\overline{k}},\cR,\cq)-H(A_i^{k]},A_j^{[k]}|\cw,\cR,\cq)\\
     &\leq H(A_i^{[k]}|W_{\overline{k}},\cR,\cq)+H(A_j^{[k]}|W_{\overline{k}},\cR,\cq),\label{eq:Ai and Aj deterministic}
\end{align}
where \eqref{eq:independence of messages fr} follows from independence of individual messages, \eqref{eq: all except Ai and Aj are deterministic} holds because the answers from all servers except $i$ and $j$ are completely determined by $W_{\overline{k}}$, $\cR$ and $\cq$, and \eqref{eq:Ai and Aj deterministic} follows since $A_i^{[k]}, A_j^{[k]}$ are completely determined by $\cw, \cR,\cq$. 
\end{Proof}

The next lemma bounds the randomness size $H(\cR)$ using the joint entropy of all answers, and holds for any SPIR setting where the common randomness is fully-replicated.

\begin{lemma}\label{lem:randomness size fr}
    For any feasible SPIR scheme with fully-replicated $\cR$, it holds that
    \begin{align}
       H(\cR)&\geq H(A_{1:N}^{[k]}|\cq)-L, \quad \forall k\in [K].
    \end{align}
\end{lemma}

\begin{Proof}
From the independence of $\cR$, $\cw$ and $\cq$ and answer generation requirement, we have
\begin{align}
    H(\cR)=&H(\cR|\cw,\cq)
    +H(A^{[k]}_{1:N}|\cw,\cR,\cq)\\
    =& H(A_{1:N}^{[k]},\cR|\cw,\cq)\\
    \geq & H(A_{1:N}^{[k]}|\cw,\cq)\label{eq: non-negativity of entropy}\\
    =& H(A_{1:N}^{[k]}|\cq) -I(\cw;A_{1:N}^{[k]}|\cq)\\
    =& H(A_{1:N}^{[k]}|\cq) - \big(I(W_k;A_{1:N}^{[k]}|\cq)+ I(W_{\overline{k}};A_{1:N}^{[k]},W_k,\cq)\big)\\
    =& H(A_{1:N}^{[k]}|\cq)-L, \label{eq: follows from reliability and database privacy}
\end{align}
where \eqref{eq: non-negativity of entropy} is due to non-negativity of entropy and \eqref{eq: follows from reliability and database privacy} follows from reliability \eqref{eq:reliability} and database privacy \eqref{eq:database_privacy2}.
\end{Proof}

Now, we motivate the proof idea by showing that $\mathscr{C}_{FR}(\mathbb{P}_3)\leq \frac{1}{2}$.

\subsubsection{Proof for $\mathbb{P}_3$}
Suppose $W_1$ is the desired message. Then, 
    \begin{align}
    L =&H(W_1|\cq)-H(W_1|A_{1:3}^{[1]},\cq)\label{eq:starting eqn p3}\\
    =&I(A^{[1]}_{1:3};W_1|\cq)\\
    =&H(A_{1:3}^{[1]}|\cq)-H(A_1^{[1]}|W_1,Q_1^{[1]},\cq)-H(A_{2:3}^{[1]}|W_1,A^{[1]}_1,\cq) \label{eq:first step generalize}\\
    \leq&H(A_{1:3}^{[1]}|\cq)-H(A_1^{[1]}|W_1,Q_1^{[1]})-H(A_{2:3}^{[1]}|W_1,A^{[1]}_1,\cq,\cR)\label{eq:follows from lemma 2}\\
    =& H(A_{1:3}^{[1]}|\cq)-H(A_1^{[2]}|W_1,Q_1^{[2]})-H(A_2^{[1]}|W_1,\cq,\cR)-H(A_3^{[1]}|W_1,A_2^{[1]},\cq,\cR) \label{eq:consequence of lemma 1}\\
    = & H(A_{1:3}^{[1]}|\cq)-H(A_1^{[1]}|Q_1^{[1]})-H(A_2^{[1]}|W_1,\cq,\cR)-H(A_3^{[1]}|W_1, A_2^{[1]},\cq,\cR) \label{eq: due to database and user privacy} \\
   \leq& H(A_{1:3}^{[1]}|\cq)-H(A_1^{[1]}|\cq)-H(A_2^{[1]}|W_1,\cq,\cR)-H(A_3^{[1]}|W_1,W_2,A_2^{[1]},\cq,\cR)\\
   =& H(A_{1:3}^{[1]}|\cq)-H(A_1^{[1]}|\cq)-H(A_2^{[2]}|W_1,\cq,\cR) ,\label{eq:sum one way p3}
\end{align}
where \eqref{eq:first step generalize} follows since $Q_1^{[1]}$ is determined from $\cq$, \eqref{eq:follows from lemma 2} follows from Lemma~\ref{lem:answer n depends only on query n} and conditioning with $\cR$,  \eqref{eq:consequence of lemma 1} holds by Lemma~\ref{lem: conseq of user priv} and since $A_1^{[1]}$ is completely determined by $(W_1,\cq,\cR)$, \eqref{eq: due to database and user privacy} follows from database and user privacy on the second term, and \eqref{eq:sum one way p3} also follows from Lemma~\ref{lem: conseq of user priv}. By interchanging the positions of $A_2^{[1]}$ and $A_3^{[1]}$ in \eqref{eq:consequence of lemma 1}, we obtain
\begin{align}
    L\leq H(A_{1:3}^{[1]}|\cq)-H(A_1^{[1]}|\cq)-H(A_3^{[2]}|W_1,\cq,\cR) .\label{eq:sum other way p3}
\end{align}
From \eqref{eq:sum one way p3} and \eqref{eq:sum other way p3}, and applying Lemma~\ref{sum of answer entropies}, we get
\begin{align}
L\leq& H(A_{1:3}^{[1]}|\cq)-H(A_1^{[1]}|\cq)-\frac{1}{2}\left(H(A_2^{[2]}|W_1,\cq,\cR)+H(A_3^{[2]}|W_1,\cq,\cR)\right) \\
\leq & H(A_{1:3}^{[1]}|\cq)-H(A_1^{[1]}|\cq)-\frac{L}{2},\label{eq:answer except 1 p3}
\end{align}
which, since $ H(A_{1:3}^{[1]}|\cq)-H(A_1^{[1]}|\cq)\leq H(A_2^{[1]}|\cq)+H(A_3^{[1]}|\cq)$, upon rearrangement gives
\begin{align}\label{eq:theta is 1}
   H(A_2^{[1]}|\cq)+H(A_3^{[1]}|\cq)\geq \frac{3L}{2}.
\end{align}
Repeating the steps in \eqref{eq:starting eqn p3}-\eqref{eq:answer except 1 p3} for $W_2$, we get
\begin{align}\label{eq:theta=2}
    H(A_1^{[2]}|\cq)+H(A_2^{[2]}|\cq)\geq \frac{3L}{2}.
\end{align}
Moreover, for $k=1,2$, by bounding $H(A_{1:3}^{[k]}|W_k,\cq)$ as
\begin{align}
   H(A_{1:3}^{[k]}|W_k,\cq)
   &= H(A_2^{[k]}|W_k,\cq)+H(A_1^{[k]},A_3^{[k]}|W_k,A_2^{[k]},\cq)\\
   &\geq H(A_2^{[k]}|\cq) + H(A_1^{[k]},A_3^{[k]}|W_1,W_2,\cR,A_2^{[k]},\cq)\\
   &= H(A_2^{[k]}|\cq) + H(A_1^{[k]},A_3^{[k]}|W_1,W_2,\cR,\cq)\\
   &=  H(A_2^{[k]}|\cq),
\end{align}
we get
\begin{align}
    H(A_1^{[k]}|\cq)+H(A_3^{[k]}|\cq)\geq L,\label{eq:sum entropy edges p3}
\end{align}
Finally,  summing \eqref{eq:sum one way p3}, \eqref{eq:sum other way p3} and \eqref{eq:sum entropy edges p3}, for $k=1,2$ by \eqref{eq:user_privacy}, we obtain
\begin{align}
2\big(H(A_1^{[k]}|\cq)+H(A_2^{[k]}|\cq)+H(A_3^{[k]}|\cq)\big)\geq 4L,\label{eq: p3 download bound}
\end{align}
which gives $\mathscr{C}_{FR}(\mathbb{P}_3)\leq \frac{1}{2}$. To determine the bound on $H(\cR)$, from Lemma~\ref{lem:randomness size fr} we have $H(\cR)\geq H(A_{1:3}^{[k]}|\cq)-L$, which satisfies 
\begin{align}
   H(A_{1:3}^{[1]}|\cq)-L&\geq H(A_1^{[k]}|\cq)+\frac{L}{2}\\
   H(A_{1:3}^{[k]}|\cq)-L&\geq H(A_2^{[k]}|\cq)\\
   H(A_{1:3}^{[2]}|\cq)-L&\geq H(A_3^{[k]}|\cq)+\frac{L}{2}.
\end{align}
From this, we obtain 
\begin{align}
    3H(\cR)\geq H(A_1^{[1]}|\cq)+H(A_2^{[k]}|\cq)+H(A_3^{[k]}|\cq)+L\geq 3L
\end{align} 
i.e., $\rho_{total}(\mathbb{P}_3)\geq 1$. Next, we generalize the proof to $\mathbb{P}_N$.

\subsubsection{Proof for Path Graphs}\label{sec:upper_bnd path}
Recall that for $\mathbb{P}_N$, $K=N-1$, and $W_n$ is replicated on servers $n$ and $n+1$, for all $n\in [N-1]$. The next lemma establishes a lower bound on the size of $A_{[1:N]\setminus\{n\}}^{[k]}$, conditioned on $A_n^{[k]}, W_k$ and $\cq$.

\begin{lemma}\label{lem:ans bound path}
For any feasible SPIR scheme on $\mathbb{P}_N$, we have
        \begin{align}
        H(A_{2:N}^{[1]}|A_1^{[1]},W_1,\cq)&\geq \frac{(N-2)L}{2}, \label{eq:answer of server 1}\\
            H(A_{1:N-1}^{[N-1]}|A_{N}^{[N-1]},W_{N-1},\cq) & \geq \frac{(N-2)L}{2}, \label{eq:answer of server N}\\
               H(A_{1:N\setminus \{n\}}^{[k]}|A_n^{[k]},W_n,\cq)&\geq \frac{(N-3)L}{2},  
               \quad k\in \{n-1,n\}, n\in \{2,\ldots,N-2\}. \label{eq:answers of non-edge servers}
        \end{align}
\end{lemma}

\begin{Proof}
The proof is based on expanding the joint entropy of answers in two orders to uncover answer pairs, each requesting the same message that they share. 

To show \eqref{eq:answer of server 1}, note that $H(A_{2:N}^{[1]}|A_1^{[1]},W_1,\cq)\geq H(A_{2:N}^{[1]}|A_1^{[1]},W_1,\cR,\cq)$ $=H(A_{2:N}^{[1]}|W_1,\cR,\cq)$, where the equality is because $H(A_1^{[k]}|W_1,\cR,\cq)=0$. Now,
\begin{align}
      H(A^{[1]}_{2:N}|W_1,\cR,\cq)
      &= \sum_{n\in [2:N]}H(A_n^{[1]}|A^{[1]}_{2:n-1},W_1,\cR,\cq)\label{eq:answer bound}\\
      &\geq \sum_{n\in [2:N]} H(A_n^{[1]}|A^{[1]}_{2:n-1},W_{1:n-1},\cR,\cq)\\
      &= \sum_{n\in [2:N-1]} H(A_n^{[n]}|W_{1:n-1},\cR,\cq)\label{eq:Nth answer is 0 and Lemma 1}\\
      &\geq \sum_{n\in [2:N-1]} H(A_n^{[n]}|W_{\overline{n}},\cR,\cq),\label{eq:answer bound one way}
\end{align}
where \eqref{eq:Nth answer is 0 and Lemma 1} holds since $H(A_N^{[1]}|W_{1:N-1},\cR,\cq)=0$ and by Lemma~\ref{lem: conseq of user priv}. Expanding \eqref{eq:answer bound} in reverse order of server indices,
\begin{align}
      H(A^{[1]}_{2:N}|A_1^{[1]},W_1,\cR,\cq)
      &= \sum_{n\in [2:N]} H(A_{n}^{[1]}|A^{[1]}_{n+1:N},W_1,\cR,\cq)\\
      &\geq \sum_{n\in [2:N]} H(A_{n}^{[1]}|A^{[1]}_{n+1:N}, W_1, W_{n:N-1},\cR,\cq)\\
      &= \sum_{n\in [3:N]} H(A_{n}^{[n-1]}| W_1, W_{n:N-1},\cR,\cq)\\
      &\geq\sum_{n\in [3:N]}H(A_{n}^{[n-1]}|W_{\overline{n-1}},\cR,\cq).\label{eq:answer bound reversed}     
\end{align}  
Summing \eqref{eq:answer bound one way} and \eqref{eq:answer bound reversed}, and using Lemma~\ref{sum of answer entropies}, we obtain
\begin{align}
   2 H(A^{[1]}_{2:N}|A_1^{[1]},W_1,\cR,\cq)
   & \geq \sum_{n=2}^{N-1} H(A_n^{[n]}|W_{\overline{n}},\cR,\cq)+H(A_{n+1}^{[n]}|W_{\overline{n}},\cR,\cq)\\
    &\geq  (N-2)L,
\end{align}
proving \eqref{eq:answer of server 1}, and \eqref{eq:answer of server N} follows the same line of argument by symmetry. To show \eqref{eq:answers of non-edge servers}, proceeding similarly, if $k=n-1$ or $k=n$, 
\begin{align}
    H(A_{1:N\setminus \{n\}}^{[k]}|A_n^{[k]},W_k,\cq) 
    \geq&    H(A_{1:N\setminus \{n\}}^{[k]}|A_n^{[k]},W_{n-1}, W_n, \cR,\cq)\label{eq:joint entropy answers} \\
    =&\sum _{n'\in [N]\setminus \{n\}}H(A_{n'}^{[k]}|A^{[k]}_{1:n'-1},W_{n-1},W_n,\cR,\cq)\\
    \geq& \sum _{n'\in [N]\setminus \{n\}}H(A_{n'}^{[k]}|W_{1:n'-1},A^{[k]}_{1:n'-1},W_{n-1}, W_n,\cR,\cq)\\
    =& \sum _{n'\in [N]\setminus \{n-1,n\}}H(A_{n'}^{[k]}|W_{1:n'-1},W_{n-1}, W_n,\cR,\cq)\\
    \geq&\sum _{n'\in [N-1]\setminus \{n-1,n\}}H(A_{n'}^{[k]}|W_{\overline{n'}},\cR,\cq)\\
    =& \sum _{n'\in [N-1]\setminus \{n-1,n\}}H(A_{n'}^{[n']}|W_{\overline{n'}},\cR,\cq)\label{eq: summing path non edge one way}.
\end{align}
Expanding \eqref{eq:joint entropy answers} in reverse order, we get
\begin{align}
    H(A_{1:N\setminus \{n\}}^{[k]}|A_n^{[k]},W_{n-1},W_n,\cR,\cq) =&\sum _{n'\in [N]\setminus \{n\}}H(A_{n'}^{[k]}|A^{[k]}_{n'+1:N},W_{n-1},W_n,\cR,\cq)\\
    \geq& \sum _{n'\in [N]\setminus \{n\}}H(A_{n'}^{[k]}|W_{n':N-1},A^{[k]}_{n'+1,N},W_{n-1}, W_n,\cR,\cq)\\
    =& \sum _{n'\in [N]\setminus \{n,n+1\}}H(A_{n'}^{[k]}|W_{n':N-1},W_{n-1}, W_n,\cR,\cq)\\
    \geq&\sum _{n'\in [2:N]\setminus \{n,n+1\}}H(A_{n'}^{[k]}|W_{\overline{n'-1}},\cR,\cq)\\
    =& \sum _{n'\in [2:N]\setminus \{n,n+1\}}H(A_{n'}^{[n'-1]}|W_{\overline{n'-1}},\cR,\cq)\label{eq: summing path non edge second way}.
\end{align}
Summing \eqref{eq: summing path non edge one way} and \eqref{eq: summing path non edge second way}, and applying Lemma~\ref{lem:sum_entropy of answers} yields for $k=n-1,n$,
\begin{align}
    H(A_{1:N\setminus \{n\}}^{[k]}|A_n^{[k]},W_k,\cR,\cq)&\geq \frac{1}{2}\left(\sum_{n'\in [N-1]\setminus \{n-1,n\}} H(A_n^{[n']}|W_{\overline{n'}},\cR,\cq)+H(A_{n+1}^{[n']}|W_{\overline{n'}},\cR,\cq)\right)\\
    &\geq \frac{(N-3)L}{2}.
\end{align}
This completes the proof of Lemma~\ref{lem:ans bound path}.
\end{Proof}
To prove the upper bound for $\mathbb{P}_N$, for $k\in [N-1]$ it follows from reliability that
\begin{align}
L&=I(A_{1:N}^{[k]};W_k|\cq)=H(A_{1:N}^{[k]}|\cq)-H(A_{1:N}^{[k]}|W_k,\cq) \label{eq:routine step start}\\
    &=H(A_{1:N}^{[k]}|\cq)-H(A_n^{[k]}|\cq)-H(A_{1:N\setminus \{n\}}^{[k]}|A_n^{[k]},W_k,\cq)\label{eq:bounding joint entropy answers}\\
    &\leq \sum_{n'\in [N]\setminus\{n\}}H(A_{n'}^{[k]}|\cq)-H(A_{1:N\setminus \{n\}}^{[k]}|A_n^{[k]},W_k,\cq).\label{eq:routine step end}
\end{align}
Bounding the second term by Lemma~\ref{lem:ans bound path} we obtain for all $k$,
\begin{align}
\sum_{n'\in [N]\setminus \{n\}} H(A_{n'}^{[k]}|\cq)\geq \begin{cases}
    \frac{NL}{2}, &n=1,N,\\
    \frac{(N-1)L}{2}, &n\in [2:N-1],
\end{cases}
\end{align}
by user privacy \eqref{eq:user_privacy}. Summing over all $n\in [N]$, we get
\begin{align}
     \sum_{n=1}^N \sum_{n'\in [N]\setminus\{n\}} H(A_{n'}^{[k]}|\cq) &\geq 2\frac{NL}{2}+(N-2)\frac{(N-1)L}{2} \\
     &= \frac{(N^2-N+2)L}{2},
\end{align}
resulting in
\begin{align}
     &(N-1)\sum_{n=1}^N H(A_n^{[k]}|\cq) \geq  \frac{(N^2-N+2)L}{2},
\end{align}
which gives 
\begin{align}
    \frac{L}{\sum_{n=1}^N H(A_n^{[k]})}\leq\frac{L}{\sum_{n=1}^N H(A_n^{[k]}|\cq)}\leq \frac{2(N-1)}{N^2-N+2}=\frac{2}{N+\frac{2}{N-1}},
\end{align}
proving the upper bound for $\mathbb{P}_N$. The minimum common randomness size, by Lemma~\ref{lem:randomness size fr} and \eqref{eq:bounding joint entropy answers} is bounded by $H(\cR)\geq H(A_n^{[k]}|\cq)+H(A_{1:N\setminus\{n\}}^{[k]}|A_n^{[k]},W_k,\cq)$. From Lemma~\ref{lem:ans bound path},
\begin{align}
     H(\cR)&\geq H(A_1^{[1]}|\cq)+\frac{(N-2)L}{2},\\
    H(\cR)&\geq H(A_2^{[1]}|\cq)+\frac{(N-3)L}{2}=H(A_2^{[2]}|\cq)+\frac{(N-3)L}{2},\\
    H(\cR)&\geq H(A_3^{[2]}|\cq)+\frac{(N-3)L}{2}=H(A_3^{[3]}|\cq)+\frac{(N-3)L}{2},\\
         \vdots& \notag\\
    H(\cR)&\geq H(A_{N-1}^{[N-2]}|\cq)+\frac{(N-3)L}{2}=H(A_{N-1}^{[N-1]}|\cq)+\frac{(N-3)L}{2},\\ 
    H(\cR)&\geq H(A_{N}^{[N-1]}|\cq)+\frac{(N-2)L}{2}.
\end{align}
Averaging over $n$ and fixing the message index to an arbitrary $k\in [K]$ for all answers by user privacy, we obtain
\begin{align}
     H(\cR)&\geq \frac{1}{N}\sum_{n=1}^NH(A_n^{[k]}|\cq)+\frac{1}{2N}(N^2-3N+2)L\\
     &\geq L\left(\frac{N-2}{2}+\frac{1}{N-1}\right),
\end{align}
which completes the proof for path graphs. 
 
\subsubsection{Proof for Cyclic Graphs}
For $\mathbb{C}_N$, $\cw_n = \{W_{n-1},W_{n}\} $ for each $ n\in [N]$ with $W_0=W_N$. The next lemma is analogous to Lemma~\ref{lem:ans bound path} for the case of $\mathbb{C}_N$.
\begin{lemma}\label{lem:answer bound cycle}
    For any feasible SPIR scheme on $\mathbb{C}_N$, and $n\in [N]$,
    \begin{align}
        H(A_{1:N\setminus \{n\}}^{[k]}|A_n^{[k]},W_n,\cq)&\geq \frac{(N-2)L}{2}, \quad k\in \{n-1,n\},
    \end{align}
    where $k=0$ implies $k=N$.
\end{lemma}

We omit the proof since it follows the same steps as that of \eqref{eq:answers of non-edge servers}. From \eqref{eq:routine step start}-\eqref{eq:routine step end} and Lemma~\ref{lem:answer bound cycle}, we obtain the bound 
\begin{align}
        &(N-1)\sum_{n=1}^N H(A_n^{[k]}|\cq) \geq N\left(L+\frac{(N-2)L}{2}\right),
\end{align}
for any achievable scheme, which yields
\begin{align}
      &\frac{L}{\sum_{n=1}^N H(A_n^{[k]})}\leq \frac{2(N-1)}{N^2}=\frac{2}{N+1+\frac{1}{N-1}}. 
\end{align}
Similarly, $H(\cR)$ is bounded as
\begin{align}
   H(\cR)&\geq \frac{1}{N}\sum_{n=1}^N H(A_n^{[k]}|\cq)+\frac{1}{2}(N-2)L \\
   &\geq \frac{L}{2}\left(N-1+\frac{1}{N-1}\right).
\end{align}
This completes the proof of Theorem~\ref{thm:upperbnd_capacity_fr}. 

\section{Extension to Multigraphs}\label{sec:ext to multigraphs}
In this section, we extend the results of Section~\ref{sec:main results} to uniform $r$-multigraphs. Our goal is to understand the variation of the SPIR metrics with $r$ under the two common randomness replication models. We begin by presenting the formal problem setting in Section~\ref{model_multigraphs}, followed by the results for the graph-replicated and fully-replicated common randomness settings in Section~\ref{sec:graph-replicated_multigraph} and Section~\ref{sec:fully-replicated_multigraph}, respectively. The extension to multigraphs is fairly straightforward, and uses similar techniques as those of simple graphs.

\subsection{Problem Setting}\label{model_multigraphs}
Following the formulation in \cite{our_journal2025}, we let $G^{(r)}$ denote the multigraph version of the simple graph $G=(V,E)$ with $V=\cs$ with constant multiplicity $r$. In $G^{(r)}=(\cs,\cw)$, each edge of $G$ is replaced with $r$ parallel edges (also referred to as multi-edges), where $r\geq 2$. To be consistent with the notations, we denote the number of edges in $G$ by $|E|=K'$, so that $K=K'r$, is the total number of edges in the multigraph, i.e., $|\cw|=K$ in the system. Let, in the storage based on $G$, the message $W_k$, $k\in [K']$ be replicated on servers $i,j\in [N]$. Then, in the corresponding multigraph-based storage, servers $i$ and $j$ share the $r$ messages, given by the set
\begin{align}\label{eq:def_w_ij}
   \mathcal{W}_{i,j}\defeq\{W_{k,1},\ldots,W_{k,r}\},
\end{align}
where $\{W_{\ell,t}$, $\ell\in [K'], t\in [r]\}$ is a relabeling of the messages $\{W_1,\ldots,W_K\}$. 

For the graph-replicated case, $\cR$ is replicated according to the simple graph $G$, while for the fully-replicated case, $\cR$ is available to all servers, irrespective of $G^{(r)}$. In the graph-replicated setting, corresponding to the message set $\cw_{i,j}$ in \eqref{eq:def_w_ij}, servers $i$ and $j$ exclusively share the randomness variable $R_k$, which we refer to as the \emph{message set-specific} randomness. The total randomness is $\cR=\{R_k, k\in [K']\}$ where $R_k$, $k\in [K']$ are i.i.d. As in Sections~\ref{sec:graph-replicated simple} and \ref{sec:fully-replicated simple}, we use $\cw_i$ and $\cR_i$ to denote the messages and randomness variables, respectively, stored at server $i$. 

Let the desired message index be $\theta=(k,\tau)$ where $k\in [K']
$ and $\tau\in [r]$. For both settings of common randomness replication, the requirements \eqref{eq:query_randomness}-\eqref{eq:reliability} follow through, by substituting $k$ with $(k,\tau)$. 

For the graph-replicated randomness setting, the database privacy requirement, for any subset $\mathcal{J}
\subseteq [K']\setminus \{k\}$ is
\begin{align}
    I(\cw_{i,j}\setminus\{W_{k,\tau}\}, W_{\mathcal{J},[r]};A_{1:N}^{[k,\tau]},Q^{[k,\tau]}_{1:N},\cR\setminus (\{R_k\}\cup R_{\mathcal{J}}),\cw\setminus (\cw_{i,j}\cup W_{\mathcal{J},[r]}),\cq)=0,\label{eq:database_privacy_multi}
\end{align}
where $W_{\mathcal{J},[r]}=\{W_{\ell,t}, \ell\in \mathcal{J}, t\in [r]\}$. 

For the fully-replicated common randomness setting, database privacy is given by
\begin{align}\label{eq:database privacy 2 multi}
    I(\cw\setminus \{W_{k,\tau}\};A_{1:N}^{[k,
\tau]},Q_{1:N}^{[k,\tau]},\cq)=0.
\end{align}
By letting $\mathcal{J}=[K']\setminus \{k\}$, we get the same requirement for \eqref{eq:database_privacy_multi} and \eqref{eq:database privacy 2 multi}, whereas for every other $\mathcal{J}$, the former
is a stronger requirement than the latter. 

In the following, we establish bounds on the SPIR capacity and the respective minimum randomness ratios for both settings.

\subsection{Graph-Replicated Common Randomness}\label{sec:graph-replicated_multigraph}
In this section, we extend the results of Section~\ref{sec:graph-replicated simple} to multigraphs. We observe that, increasing the replicated message set size from $1$ to $r>1$, does not affect the results, i.e., the SPIR metrics of $G^{(r)}$ are independent of $r$.

\subsubsection{General Scheme and Converse}

\begin{theorem}
    The SPIR capacity of a general $r$-multigraph $G^{(r)}$ is bounded as
    \begin{align}
        \mathscr{C}(G^{(r)})\geq \frac{1}{N},
    \end{align} 
    provided that the randomness ratio $\rho^*\leq \rho = 1$.
\end{theorem}

\begin{Proof}
The proof follows by applying the scheme in  Section~\ref{subsec:gen_scheme_graphs} to multigraph-based storage, through the following generalization. Let each message and randomness consist of $L=1$ symbol. As before, the user chooses $K$ random symbols $h_{\ell,t}, \ell\in [K'], t\in [r]$ and forms the vectors $\tilde{\bm{h}}_t=[h_{1,t},\ldots,h_{K',t}]$ for each $t\in [r]$. Define $\bm{H}^{(r)}$ as the $N\times K$ query matrix
\begin{align}
    \bm{H}^{(r)}=
    \begin{bmatrix}
        \bar{I}(G)\cdot \text{diag}(\tilde{\bm{h}}_1),& \bar{I}(G)\cdot \text{diag}(\tilde{\bm{h}}_2),& \ldots, & \bar{I}(G)\cdot \text{diag}(\tilde{\bm{h}}_r)
    \end{bmatrix},
\end{align}
where $\bar{I}(G)$ is the $N\times K'$ signed incidence matrix of $G$ and $\bm{H}^{(r)}$ is $N\times K$. Similarly, we prepare the $r|\mathcal{F}_n|\times 1$ message vector $\bm{W}_n$ for server $n$ as
\begin{align}
    \bm{W}_n=\begin{bmatrix}
        \begin{bmatrix}W_{\ell,1}\end{bmatrix}_{\ell\in \mathcal{F}_n}\\
        \vdots\\
        \begin{bmatrix}W_{\ell,r}\end{bmatrix}_{\ell\in \mathcal{F}_n}
    \end{bmatrix},
\end{align}
where $\mathcal{F}_n:=(\ell:W_{\ell,t}\in \mathcal{W}_n, \forall t\in [r])$, with the indices arranged in an ascending order, and $\begin{bmatrix}W_{\ell,t}\end{bmatrix}_{\ell\in \mathcal{F}_n}$ is the vertical concatenation of $W_{\ell,t}$ for all $\ell\in \mathcal{F}_n$, with $|\mathcal{F}_n|=\deg(n)$. 

Let the desired message be $W_{k,\tau}$, where $k\in [K']$ and $\tau\in [r]$. Assume, without loss of generality, that $W_{k,\tau}\in \cw_{i,j}$, where $W_{k,\tau}$ is the $m$th message of the vector $\bm{W}_j$. Let $\bm{h}_n$ be the $n$th row of $\bm{H}^{(r)}$, after discarding the zeros, i.e., $\bm{h}_n$ is a row vector of length $r\cdot \text{deg}(n)$. To server $n$, the query sent is
\begin{align}
    Q_n^{[k,\tau]}&=
    \begin{cases}
        \bm{h}_n^\top, & n\in [N]\setminus \{j\},\\
        \bm{h}_n^\top+\bm{e}_m, &n=j,
    \end{cases}
\end{align}
and the answer sent in response is
\begin{align}
    A_n^{[k,\tau]}=Q_n^{[k,\tau]\top} \bm{W}_n + \sum_{\ell\in \mathcal{F}_n}\bar{I}(G)(n,\ell)\cdot R_{\ell}.
\end{align}
The decoding at the user is done by computing the sum of all $A_n^{[k,\tau]}$, i.e., 
\begin{align}
    \sum_{n\in [N]}& \left (\sum_{t=1}^r\left( \sum_{\ell\in \mathcal{F}_n} \bar{I}(G)(n,\ell) h_{\ell,t} W_{\ell,t}\right) + \sum_{\ell\in \mathcal{F}_n}R_{\ell}\right)+W_{k,\tau} \nonumber\\
    &= \left(\sum_{\ell\in \mathcal{F}_n}  \bigg(\sum_{t=1}^rh_{\ell,t} W_{\ell,t}\bigg) + R_{\ell}\right)\underbrace{\left( \sum_{n\in [N]} \bar{I}(G) (n,\ell) \right)}_{=0} + W_{k,\tau}\\
    &=W_{k,\tau},    
\end{align}
the user decodes $W_{k,\tau}$. Hence, reliability is ensured. The user downloads a total of $N$ answer symbols to retrieve $L=1$ message symbol, resulting in the rate $\frac{1}{N}$. 

For user privacy, note that each server receives a uniformly distributed vector of length $r\cdot \deg(n)$ with symbols from $\mathbb{F}_q$ as query, irrespective of the desired message.
Database privacy is preserved due to one-time padding of each message set in the answers. Specifically, no information on any subset of $\{W_{\ell,[r]},\ell\in [L]\setminus \{k\}\}$ is revealed to the user, since the corresponding subset of $\{R_\ell, \ell\in [L]\setminus \{k\}\}$ is unavailable to them. Similarly, within the desired message set, the messages $\{W_{k,t}, t \in [r]\setminus \{\tau\}\}$ are protected since $R_k$ is not known to the user.   
\end{Proof}

\begin{theorem}\label{thm:rho bound multigraph}
    For an SPIR scheme on $G^{(r)}$ to be feasible, $\rho\geq 1$, i.e., $\rho^*=1$.
\end{theorem}

The proof of Theorem~\ref{thm:rho bound multigraph} follows the same arguments as the simple graph ($r=1$) case. Observation \eqref{eq:R_k_deterministic} and Lemmas~\ref{lem:answers_independent_of_index} and \ref{lem:answer_indep_randomness_given_query} hold for any index $(k,\tau)$. 

The next lemma is the multigraph version of Lemma~\ref{lem:sum_entropy of answers}.

\begin{lemma}\label{lem:sum of answer entropies multi}
    For any $W_{k,\tau}$ replicated on servers $i$ and $j$, it holds that
    \begin{align}
        H(A_i^{[k,\tau]}|\cR\setminus\{R_k\}, \cw\setminus \cw_{i,j},\cq)+H(A_j^{[k,\tau]}|\cR\setminus \{R_k\}, \cw\setminus \cw_{i,j},\cq)\geq (1+\rho)L.
    \end{align}  
\end{lemma}

\begin{Proof}
Let $\theta=(k,\tau)$ and assume that $W_{k,\tau}\in \cw_{i,j}$. Let $\cw'\defeq \cw\setminus \cw_{i,j}$ and $R_{\overline{k}} \defeq \cR\setminus \{R_k\}$. Then, by expanding $I(A_{1:N}^{[k,\tau]};\cw_{i,j}|R_{\overline{k}}, \cw',\cq)$ in one way, we have
\begin{align}
    I(&A_{1:N}^{[k,\tau]};\cw_{i,j}|R_{\overline{k}}, \cw',\cq)\nonumber\\
    &=H(\cw_{i,j}|R_{\overline{k}}, \cw',\cq)-H(\cw_{i,j}|A_{1:N}^{[k,\tau]},R_{\overline{k}}, \cw',\cq)\label{eq:mutual_info term in first way}\\
    &=rL-\big(H(W_{k,\tau}|A_{1:N}^{[k,\tau]},R_{\overline{k}}, \cw',\cq)+H(\cw_{i,j}\setminus \{W_{k,\tau}\}|A_{1:N}^{[k,\tau]},R_{\overline{k}}, \cw',W_{k,\tau},\cq)\big)\\
    &= rL - H(\cw_{i,j}\setminus \{W_{k,\tau}\}|A_{1:N}^{[k,\tau]},R_{\overline{k}}, \cw',\cq) \label{eq: follows from reliability on terms}\\
    &=rL-(r-1)L \label{eq: follows from db privacy}\\
    &=L,\label{eq:mutual info one way}
\end{align}
where \eqref{eq: follows from reliability on terms} follows since $W_{k,\tau}$ is deterministic given $A_{1:N}^{[k,\tau]}$ and $\cq$, and \eqref{eq: follows from db privacy} follows from database privacy \eqref{eq:database_privacy_multi} with $\mathcal{J}=\emptyset$. On the other hand, by following the steps in the proof of Lemma~\ref{lem:sum_entropy of answers}, we obtain
\begin{align}
    I(A_{1:N}^{[k,\tau]};\cw_{i,j}|R_{\overline{k}}, \cw',\cq)
    \leq H(A_i^{[k,\tau]}|R_{\overline{k}}, \cw',\cq)+H(A_j^{[k,\tau]}|R_{\overline{k}}, \cw',\cq)-\rho L.\label{eq:mutual info second way}
\end{align}
From \eqref{eq:mutual info one way} and \eqref{eq:mutual info second way}, we obtain the inequality. \end{Proof}

\begin{Proof} \textbf{(of Theorem~\ref{thm:rho bound multigraph})} 
Let $\theta = (k',\tau)$ with $\cw_{i',j'}=W_{k',[r]}$ where $\{i',j'\}\neq \{i,j\}$. The database privacy \eqref{eq:database_privacy_multi} with $\mathcal{J}=\{k\}$ where $k\neq k'$ gives 
\begin{align}
    0 =& I(\cw_{i',j'}\setminus \{W_{k',\tau}\},W_{k,[r]};A_1^{[k',\tau]},\ldots,A_N^{[k',\tau]},R_{\overline{k}}\setminus \{R_{k'}\},\cw\setminus (\cw_{i',j'}\cup W_{k,[r]}),\cq)\\
    =& I(\cw_{i',j'}\setminus \{W_{k',\tau}\};A_1^{[k',\tau]},\ldots,A_N^{[k',\tau]},R_{\overline{k}}\setminus \{R_{k'}\},\cw\setminus (\cw_{i',j'}\cup W_{k,[r]}),\cq) \notag \\
    &+ I(W_{k,[r]};A_1^{[k',\tau]},\ldots,A_N^{[k',\tau]},R_{\overline{k}}\setminus \{R_{k'}\},\cw\setminus W_{k,[r]},\cq)\notag \\
    &+ I( W_{k,[r]};R_{k'}|A_1^{[k',\tau]},\ldots,A_N^{[k',\tau]},R_{\overline{k}}\setminus \{R_{k'}\},\cw\setminus W_{k,[r]},\cq), \label{eq: expanding mutual info}
\end{align}
where the first term in \eqref{eq: expanding mutual info} is lower than $I(\cw_{i',j'}\setminus \{W_{k',\tau}\};A_1^{[k',\tau]},\ldots,A_N^{[k',\tau]},\cR \setminus \{R_{k'}\},\cw\setminus \cw_{i',j'},\cq)$ which is zero by database privacy and the third term added to \eqref{eq: expanding mutual info} is zero since
    \begin{align} H(R_{k'}|A_1^{[k',\tau]},\ldots,A_N^{[k',\tau]},R_{\overline{k}}\setminus \{R_{k'}\},\cw\setminus W_{k,[r]},\cq)=0, \label{eq:follows from the obsercation} 
\end{align} 
where \eqref{eq:follows from the obsercation} follows by noting that at least one of servers $i'$ or $j'$ stores a subset of $\cw\setminus W_{k,[r]}$ and consequently one can decode $R_{k'}$ given the answer from the respective server, $R_{\overline{k}}\setminus \{R_{k'}\}$, $\cw\setminus W_{k,[r]}$ and $\cq$. Rewriting $W_{k,[r]}$ as $\cw'$ in \eqref{eq: expanding mutual info}, we obtain
\begin{align}
    0
    =& I(W_{k,[r]};A_1^{[k',\tau]},\ldots,A_N^{[k',\tau]}|R_{\overline{k}},\cw',\cq)\\
    \geq & I(W_{k,[r]};A_i^{[k',\tau]}|R_{\overline{k}},\cw',\cq)\\
    =&H(A_i^{[k,\tau]}|\cw',R_{\overline{k}},Q_i^{[k,\tau]})-H(A_i^{[k,\tau]}|\cw,\cR,Q_i^{[k,\tau]})-I(R_k;A_i^{[k,\tau]}|\cw,R_{\overline{k}},Q_i^{[k,\tau]})\\
    = & H(A_i^{[k,\tau]}|\cw',R_{\overline{k}},\cq)-H(R_k).
\end{align}Similarly, we get
\begin{align}
    0 \geq  H(A_j^{[k,\tau]}|\cw',R_{\overline{k}},\cq)-H(R_k),
\end{align}
which combined with Lemma~\ref{lem:sum of answer entropies multi} gives $H(R_k) = \rho L \geq L$, completing the proof. 
\end{Proof}

\begin{theorem}
    For the multigraph extensions $\mathbb{P}_N^{(r)}$ and $G^{(r)}$ where $G$ is $d$-regular graph, the SPIR capacity is $\frac{1}{N}$.    
\end{theorem}

It suffices to show that the capacity upper bound is $\frac{1}{N}$. For $d$-regular multigraphs, this follows directly from Theorem~\ref{thm:rho bound multigraph} and  Lemma~\ref{lem:sum of answer entropies multi}. For $\mathbb{P}_N^{(r)}$, we separate into $N$ even and odd, and complete the proof using the multigraph version of Lemma~\ref{lem:single_server_answer_entropy}, as stated next. 

\begin{lemma}\label{lem:single_server_answer_entropy_multi}
    For server $S$ with degree of $G=\delta$, let the neighbor set be $\cn(S)=\{S_1,\ldots,S_\delta\}$. Then, the following bound holds
    \begin{align}
        H(A_{S}^{[k,\tau]}|\cq)\geq \sum_{i=1}^{\delta}\max\left\{0,2L-\sum_{j=i}^\delta H(A_{S_j}^{[k,\tau]}|\cq)\right\}.
    \end{align}
\end{lemma}

\begin{Proof} 
Let $W_{i, [r]}$ and $R_i$, be the message set and randomness variable shared by servers $S$ and $S_i$, respectively, and  $\cw^{c}:=\cw\setminus \{\cup_{i=1}^{\delta}W_{i, [r]}\}$ and $\cR^c=\cR\setminus \{\cup_{i=1}^{\delta}R_i\}$. Then, 
\begin{align}
    H(&A_{S}^{[k,\tau]}|\cq)\notag\\
    =&I(A_{S}^{[k,\tau]};W_{1,[r]},\ldots, W_{\delta,[r]},R_{[\delta]}|\cq,\cw^c,\cR^c)\\
    =&\sum_{i=1}^{\delta} I(A_{S}^{[i,\tau]};W_{i,[r]},R_{i}|W_{1,[r]},\ldots, W_{\delta,[r]},R_{[i-1]},\cq,\cw^c,\cR^c)\\
    =&\sum_{i=1}^{\delta} H(W_{i,[r]},R_{i}|W_{1,[r]},\ldots, W_{\delta,[r]},R_{[i-1]},\cq,\cw^c,\cR^c) \notag \\
    &-\sum_{i=1}^{\delta} H(W_{i,[r]},R_{i}|A_S^{[i,\tau]},W_{1,[r]},\ldots, W_{\delta,[r]},R_{[i-1]},\cq,\cw^c,\cR^c)\\
    \geq& \sum_{i=1}^{\delta}\left((r+1)L-H(W_{i,[r]},R_{i}|A_S^{[i,\tau]},W_{1,[r]},\ldots, W_{i-1,[r]},R_{[i-1]},\cq,\cw^c,\cR^c)\right),\label{eq:bounding_first term}
\end{align}
where \eqref{eq:bounding_first term} holds because $H(R_{i}|W_{i,[r]},R_{[i-1]},\cq,\cw^c,\cR^c)\geq L.$
Now, we derive an upper bound on the second term in the summand for each $i\in [\delta]$. We use the notation $A^{[i,\tau]}_{i:\delta}\defeq\{A_{S_i}^{[i,\tau]},\ldots,A_{S_{\delta}}^{[i,\tau]}\}$. Following the steps in Lemma~\ref{lem:single_server_answer_entropy}, we obtain for each $i\in [\delta]$ that
\begin{align}
    H(W_{i,[r]}, &R_{i}|A_{S}^{[i,\tau]},W_{1,[r]},\ldots, W_{i-1,[r]},R_{[i-1]},\cq,\cw^c,\cR^c)\notag \\
    =& H(W_{i,[r]}|A_{S}^{[i,\tau]},W_{1,[r]},\ldots, W_{i-1,[r]},R_{[i-1]},\cq,\cw^c,\cR^c,A_{i:\delta}^{[i,\tau]})\notag \\
    &+H(R_i|A_{S}^{[i,\tau]},W_{1,[r]},\ldots, W_{i-1,[r]},\cq,\cw^c,\cR^c,A_{i:\delta}^{[i,\tau]})\notag\\
    &+I(W_{i,[r]}, R_{i};A_{i:\delta}^{[i,\tau]}|A_{S}^{[i,\tau]},W_{1,[r]},\ldots, W_{i-1,[r]},\cq,\cw^c,\cR^c)\\
    =& H(W_{i,{[r]\setminus \{\tau\}}}|A_{S}^{[i,\tau]},W_{i-1,[r]},R_{[i-1]},\cq,\cw^c,\cR^c,A_{i:\delta}^{[i,\tau]})\nonumber\\
    &+I(W_{i,[r]}, R_{i};A_{i:\delta}^{[i,\tau]}|A_{S}^{[i,\tau]},W_{1,[r]},\ldots, W_{i-1,[r]},\cq,\cw^c,\cR^c) \label{eq: due to db privacy}\\
    \leq& (r-1)L + \sum_{j=i}^{\delta}H(A_{S_j}^{[k,\tau]}|\cq)\label{eq:bound each second term},
\end{align}
where the first term in \eqref{eq: due to db privacy} is $(r-1)L$ due to database privacy. Combining \eqref{eq:bounding_first term} and \eqref{eq:bound each second term} completes the proof of the lemma.
\end{Proof}

\subsubsection{SPIR from PIR Schemes}\label{sec:achievable gr multi srp}
Echoing the PIR-induced construction of Section~\ref{sec:spir from pir gr}, we present an alternative SPIR scheme derived from PIR scheme on multigraphs. Interestingly, the rate achieved through this scheme is also $\frac{R_{PIR}(G)}{2}$, independent of the multiplicity $r$. 

\begin{theorem}
    For a PIR scheme $T$ on $G$, which satisfies the SRP, let the rate be $R_{PIR}(G)$. Then, there exists an SPIR scheme for $G^{(r)}$, with rate $R(G)=\frac{R_{PIR}(G)}{2}$ and $\rho = 1$.
\end{theorem}

Recall that, $L'$ denotes the number of symbols per message and $D'$ denotes the number of downloaded symbols, corresponding to the PIR scheme $T$ on $G$. Moreover, since $T$ admits SRP, there exists a PIR scheme $T^{(r)}$ on $G^{(r)}$, given by the construction in \cite{our_journal2025}. Then, in the SPIR scheme on $G^{(r)}$, we let $L=2^{r-1}L'$ and $\rho = 1$. 

The following example on multipath graph with $r=2$ illustrates the main idea of the algorithm.

\begin{example}[Multipath $\mathbb{P}_3^{(2)}$]
We have $K=2\cdot 2=4$ and $L=2^{2-1}\cdot 2=4$, since $K'=L'=2$ by Example~\ref{ex:scheme2_path}. Let us denote the message symbols after permutation as $W_{1,1}=(a_1,a_2,a_3,a_4)$, $W_{2,1}=(b_1,b_2,b_3,b_4)$, $W_{1,2}=(c_1,c_2,c_3,c_4)$ and $W_{2,2}=(d_1,d_2,d_3,d_4)$. Similarly, the common randomness symbols corresponding to the multi-edges upon independent permutations are $R_1=(s_{1,1},\ldots,s_{1,4})$ and $R_2=(s_{2,1},\ldots,s_{2,4})$. The SPIR answer table for all $\theta$ is shown in Table~\ref{tab:spir_path_r=2}, resulting in the rate $R(\mathbb{P}_3^{(2)})=\frac{1}{3}=R(\mathbb{P}_3)$.
\end{example}

\begin{table}[htbp]
    \centering
    \begin{tabular}{|c|c|c|c|}
    \hline
    \multirow{4}{*}{$\theta=(1,1)$} & database 1 & database 2 & database 3\\
    \hline
    &$s_{1,1}$ & $s_{1,2}+s_{2,2}$ & $s_{2,2}$\\
   &  $a_1+s_{1,2}$  & $a_2+b_2+s_{1,1}+s_{2,1}$ & $b_2+s_{2,1}$ \\
    & $c_1+s_{1,3}$ & $c_2+d_2+s_{1,4}+s_{2,3}$ & $d_2+s_{2,3}$\\
    & $a_4+c_2+s_{1,4}$ & $a_3+b_4+c_1+d_4+s_{1,3}+s_{2,4}$ & $b_4+d_4+s_{2,4}$\\
    \hline
    \multirow{4}{*}{$\theta=(1,2)$} & $s_{1,1}$ & $s_{1,2}+s_{2,2}$ & $s_{2,2}$\\
    & $a_1+s_{1,3}$  & $a_2+b_2+s_{1,4}+s_{2,3}$ & $b_2+s_{2,3}$ \\
    & $c_1+s_{1,2}$ & $c_2+d_2+s_{1,1}+s_{2,1}$ & $d_2+s_{2,1}$\\
    & $a_2+c_4+s_{1,4}$ & $a_1+b_4+c_3+d_4+s_{1,3}+s_{2,4}$ & $b_4+d_4+s_{2,4}$\\
    \hline
    \multirow{4}{*}{$\theta=(2,1)$} & $s_{1,1}$ & $s_{1,1}+s_{2,1}$ & $s_{2,2}$\\
    & $a_1+s_{1,2}$  & $a_1+b_1+s_{1,2}+s_{2,2}$ & $b_2+s_{2,1}$ \\
    & $c_1+s_{1,3}$ & $c_1+d_1+s_{1,3}+s_{2,3}$ & $d_2+s_{2,4}$\\
    & $a_4+c_4+s_{1,4}$ & $a_4+b_4+c_4+d_2+s_{1,4}+s_{2,4}$ & $b_3+d_1+s_{2,3}$\\
    \hline
    \multirow{4}{*}{$\theta=(2,2)$} & $s_{1,1}$ & $s_{1,1}+s_{2,1}$ & $s_{2,2}$\\
    & $a_1+s_{1,3}$  & $a_1+b_1+s_{1,3}+s_{2,3}$ & $b_2+s_{2,4}$ \\
    & $c_1+s_{1,2}$ & $c_1+d_1+s_{1,2}+s_{2,2}$ & $d_2+s_{2,1}$\\
    & $a_4+c_4+s_{1,4}$ & $a_4+b_2+c_4+d_4+s_{1,4}+s_{2,4}$ & $b_1+d_3+s_{2,3}$\\
    \hline
    \end{tabular}
    \caption{Answer table for the SPIR scheme on $\mathbb{P}_3^{(2)}$}
    \label{tab:spir_path_r=2}
\end{table}

For the general scheme, let $\theta=(k,\tau)$ denote the desired message index. It is clear from the example that the first line of answers in each row of Table~\ref{tab:spir_path_r=2} is identical to that of the algorithm for simple graphs, i.e., applying $T$ on $\cR$ to retrieve $L'$ symbols of $R_k$. The steps of our algorithm to retrieve $W_{k,\tau}$ are as follows. 
 
\textit{Index Permutation:} The user permutes the symbol indices of $W_{\ell,t}, \ell\in[K'],t\in[r]$ using $K$ random and independent permutations. Independent of the permutations on the message symbols, the user assigns $K'$ random and independent permutations to the symbol indices of $R_\ell, \, \ell\in [K']$. Let the permuted randomness symbols be denoted as $R_\ell = (s_{\ell,1},\ldots,s_{\ell,L})$ for each $\ell\in [K']$.

\textit{Query Generation:} The user generates two types of queries, (i) queries of $T$ on $\cR$, repeated $2^r$ times, each time to retrieve $L'$ (not necessarily distinct) symbols of $R_k$, (ii) queries of $T^{(r)}$ on $\cw$, with the desired index $\theta=(k,\tau)$, as specified by the construction in \cite{our_journal2025}. The queries of type (i) can be further partitioned into one stand-alone query (corresponding to the first line of answers in Table~\ref{tab:spir_path_r=2}), and remaining $(2^{r}-1)$ queries. The stand-alone query retrieves the \emph{first} $L'$ permuted symbols of $R_k$. Each of the latter  $(2^{r}-1)$ queries is \emph{associated with} a query of type (ii), as shown in Table~\ref{tab:type i) and ii) query association}. To illustrate how the queries of type (i) and (ii) are associated, we briefly recall the construction of $T^{(r)}$.

The scheme $T^{(r)}$ proceeds in $r$ stages. In the first stage, scheme $T$ is applied $r$ times to the message subsets of $\cw$, 
\begin{align}\label{eq:stage 1 msg set}
    \{W_{\ell,1} \,,\ell\in[K']\},\{W_{\ell,2} \,,\ell\in[K']\},\ldots, \{W_{\ell,r} \,,\ell\in[K']\} ,   
\end{align}
to retrieve $L'$ symbols, each of $W_{k,1}, W_{k,2},\ldots,W_{k,r}$, respectively. The retrieved symbols of $W_{k,t}, t\in[r]\setminus \{\tau\}$ act as side information for the answers in stage 2. In the subsequent stages $2\leq s\leq r$, consider the $s$-sized subsets $A\subseteq [r]$. For every $A$, scheme $T$ is applied to the sums of messages,
\begin{align}\label{eq:stage s msg set}
    \left\{  \sum_{m\in A} W_{1,m}, \sum_{m\in A} W_{2,m}, \ldots, \sum_{m\in A} W_{K',m}\right\},
\end{align} 
to retrieve $L'$ symbols of the sum $\sum_{m\in A} W_{k,m}$. In each stage, $1\leq s\leq r-1$, the retrieved sums without symbols of $W_{k,\tau}$ act as interference for the next stage. They are designed to be canceled from the sums containing symbols of $W_{k,\tau}$ in stage $s+1$. This proceeds till $s=r$ and corresponds to a total of $\sum_{s=1}^r \binom{r}{s}=2^{r}-1$ queries of $T$, across all stages. 

Now, we go back to the description of SPIR queries. For the stand-alone query of type (i) sent to server $n\in [N]$, let $P_{\ell,n}^1 \subsetneq [L']$ denote the queried symbol indices of $R_\ell$, for each $\ell\in \mathcal{F}_n$. By the SRP of $T$, each $ |P_{\ell,n}^1|=\frac{L'}{2}$ . Given this, the mapping of the remaining queries of type (i) to those of type (ii) in all stages is as follows. In the first stage, corresponding to the type (ii) query set that retrieves symbols of $W_{k,\tau}$, we retrieve the first $L'$ permuted symbols of $R_k$. More specifically, we assign a bijective mapping $\phi^1:[L'] \to [L']$ such that for any server $n$, and $\ell\in \mathcal{F}_n$, every $\mu\in P_{\ell,n}^1$ maps into $\phi^1(\mu)\in [L']\setminus P_{\ell,n}^1$. Thus, the associated query of type (i) sent to server $n$ requests for the randomness symbol indices $\phi^1(P^1_{\ell,n})$. Next, corresponding to each of the $(r-1)$ queries of $T$ that retrieve the interference symbols of $W_{k,t}, t\in [r]\setminus \{\tau\}$, the user retrieves new $L'$ permuted symbols of $\cR$. Specifically, the query to retrieve $W_{k,t}$, is mapped to the query to retrieve the permuted symbols $(x-1)L'+1:xL'$ of $R_k$,  where
\begin{align}\label{define_x}
    x=\begin{cases}
       t+1, & t\in [1:\tau-1]\\
        t,& t\in [\tau+1:r].
    \end{cases}
\end{align}
Further, the symbol indices of $\cR$ queried from server $n$ are given by
\begin{align}
   P_{\ell,n}^x=\{\mu+(x-1)L':\mu\in P_{\ell,n}^1\},  
\end{align}
where $x\in[2:r]$ is defined in \eqref{define_x}. This completes the query mapping for stage 1. 

In stage 2, corresponding to the $(r-1)$ queries of $T$ for the retrieval of $\{W_{k,\tau}+W_{k,t}, t\in [r]\setminus\{\tau\}\}$, we assign queries to retrieve the permuted symbols $[L'+1:rL']$ of $R_k$. More specifically, for each $t$, we assign $(r-1)$ bijective mappings $\phi^2,\ldots, \phi^r$, $\phi^x:[(x-1)L'+1:xL']\to [(x-1)L'+1:xL']$ such that, for every $n$, $\mu\in P_{\ell,n}^x$ maps into $\phi^x(\mu)\in [(x-1)L'+1:xL']\setminus P_{\ell,n}^x$, for $\ell\in\mathcal{F}_n$. For the remaining $\binom{r-1}{2}$ queries for the retrieval of $\{W_{k,t}+W_{k,t'}, \{t,t'\}\in [r]\setminus \{\tau\}\}$, we assign queries to retrieve the next $\binom{r-1}{2}$ symbols of $R_k$. To do this, we arrange all the $2$-subsets $\{t,t'\}\subset [r]\setminus \{\tau\}$ in a lexicographic order. Let $x$ denote the position of a pair $\{t,t'\}$ under this ordering. Then, corresponding to the retrieval of the $x$th sum $W_{k,t}+W_{k,t'}$, queries of type (i) to retrieve the permuted symbols $rL'+[(x-1)L'+1:xL']$ of $R_k$ are assigned. This completes the mapping for stage 2. 

In a general stage $s$, the queries of type (i) depend on whether symbols of $W_{k,\tau}$ are present in the sum of messages retrieved through the queries of type (ii). If present, we send queries to retrieve the $\binom{r-1}{s-1}L'$ symbols of $R_k$, retrieved in stage $s-1$ by mapping the queried indices through bijections. The bijections ensure that the same randomness symbol is not queried from any given server, in both stages $s-1$ and $s$. If $W_{k,\tau}$ does not appear in the retrieved sum in the type (ii) queries, we assign type (i) queries to retrieve the next $\binom{r-1}{s}L'$ symbols of $R_k$. In particular, corresponding to the retrieval of the $x$th sum of $s$ messages (according to the lexicographic ordering of $s$-sized subsets of $[r]\setminus \tau$), the assigned type (i) queries  retrieve the symbols of indices $\sum_{i=0}^{s-1}\binom{r-1}{i}L'+[(x-1)L'+1:xL']$ of $R_k$. 

\begin{table}[htbp]
    \centering
    \begin{tabular}{p{2cm} p{6.0cm} p{7.0cm}}
    \hline
       stage & message sums retrieved  by type (ii) queries & symbol indices of $R_k$ retrieved by type (i) queries \\
        \hline
       \hline
       - & - & $1:L'$ \\
       \hline
       \multirow{2}{*}{1} & $W_{k,\tau}$ & $1:L'$  \\
       & $W_{k,t}, t\in [r]\setminus \{\tau\}$ & $L'+1:rL'$ \\
       \hline
       \multirow{2}{*}{2} & $W_{k,\tau}+W_{k,t}, t\in [r]\setminus \{\tau\}$ & $L'+1:rL'$  \\
       &  $W_{k,t}+W_{k,t'}, t,t'\in [r]\setminus \{\tau\}$ & $rL'+1:\left(r+\binom{r-1}{2}\right)L'$\\
       \hline
       $\cdots$ & $\cdots$ & $\cdots$ \\
       \hline
       \multirow{2}{*}{$s$} & $\sum_{m\in A:\tau\in A}W_{k,m}, A\subseteq [r], |A|=s$ & $\sum_{i=0}^{s-2}\binom{r-1}{i}L'+1:\sum_{i=0}^{s-1}\binom{r-1}{i}L'$ \\
       & $\sum_{m\in A:  \tau \notin A}W_{k,m}, A\subseteq [r], |A|=s$ 
       & $\sum_{i=0}^{s-1} \binom{r-1}{i}L'+1:\sum_{i=0}^{s}\binom{r-1}{i}L'$\\
       \hline
       $\cdots$ & $\cdots$ & $\cdots$ \\
       \hline
       $r$ & $\sum_{t=1}^r W_{k,t}$ & $(2^{r-1}-1)L'+1:2^{r-1}L'$ \\
       \hline
    \end{tabular}
    \caption{Association of type (i) and type (ii) queries.}
    \label{tab:type i) and ii) query association}
\end{table}

\textit{Answer Formation:} In response, the servers respond with two sets of answers. The first set is the response to the stand-alone query of type (i), the servers return answers of $T$ applied to retrieve $(s_{k,1},\ldots,s_{k,L'})$. The second set is the answers of type (ii), added to the respective answers of type (i) queries, associated in the manner described earlier. 

The user retrieves $W_{k,\tau}$ because of the reliability of $T^{(r)}$ through successive interference cancellation across stages. In particular, from the answers of stage $s\in [r]$, the user recovers the desired message symbols (upper half of each row of Table \ref{tab:type i) and ii) query association}). This is done by canceling out the respective masked interference symbols, retrieved in the previous stage.

The construction follows the query structure of $T^{(r)}$ that follows user-privacy in the PIR setting. Moreover, the association of type (i) and type (ii) queries does not reveal any information on the desired message index. This is because the private, independent permutation applied to the symbols of $\cR$ hides the actual indices that are queried. 

Database privacy is preserved because of one-time padding of the interference symbols by $R_{\ell}, \ell\in [K']\setminus \{k\}$. The retrieved symbols of $R_k$ are added solely to the symbols of $W_{k,\tau}$, while those added to $W_{k,t},t\neq \tau$ and their sums with $W_{k,\tau}$ are unavailable to the user. Moreover, the randomness symbols added to the interference symbols are distinct from those that are queried through the stand-alone query of type (i). 

The rate of the SPIR scheme on $G^{(r)}$, in terms of the PIR rate on $G$ is given by
\begin{align}
    R(G^{(r)})&=\frac{2^{r-1}L'}{D'(2^r-1)+D'}\\
    &=\frac{R_{PIR}(G)}{2},
\end{align}
which is equal to $R(G)$. The SPIR rate, with graph-replicated randomness, is once again independent of the message multiplicity $r$. 

\subsection{Fully-Replicated Common Randomness}\label{sec:fully-replicated_multigraph}
In this section, we extend the results of Section~\ref{sec:fully-replicated simple} to multigraphs. On replicating $\cR$ at all servers, the SPIR performance naturally improves compared to that in Section~\ref{sec:graph-replicated_multigraph}. We demonstrate this by establishing capacity bounds in the following results. Clearly distinct from the graph-replicated randomness case, the bounds on the SPIR metrics depend on $r$. Moreover, with $r=1$, we recover the bounds for the corresponding simple graphs.

\begin{theorem}\label{thm:achievable_multigraphs fr}
    For any PIR scheme on $G$ with $N$ vertices and $K'$ edges, that satisfies SRP, there exists a corresponding SPIR scheme on the multigraph $G^{(r)}$ with the rate
    \begin{align}\label{eq:ach rate multigraph fr}
        {R}_{FR}(G^{(r)})= \frac{2^{r-1}L'x}{(2^r-1)D'x+Ny},
    \end{align}
    and total randomness ratio
    \begin{align}
        \rho_{total}(G^{(r)}) = \frac{Ny}{2^{r-1}xL'}-\frac{K'+1}{2^r}+K',
    \end{align}
    where $x$, $y$, $L'$ and $D'$ have the same definitions as in Theorem~\ref{thm:spir scheme fr} with respect to the starting PIR scheme.
\end{theorem}

As a direct consequence of this, we have the following bounds
\begin{align}
    \mathscr{C}_{FR}(\mathbb{P}_N^{(r)})&\geq R_{FR}(\mathbb{P}_N^{(r)})= \frac{2}{N(2-2^{1-r})+\frac{N}{N-1}2^{1-r}}, \\
    \mathscr{C}_{FR}(\mathbb{C}_N^{(r)})&\geq R_{FR}(\mathbb{C}_N^{(r)})=\frac{2}{(N+1)(2-2^{1-r})+\frac{N}{N-1}2^{1-r}},
\end{align}
that are achievable with total randomness ratios,
\begin{align}
    \rho^*_{total}(\mathbb{P}_N^{(r)})\leq \rho_{total}(\mathbb{P}_N^{(r)})&=\frac{1}{R_{FR}(\mathbb{P}_N^{(r)})}-1\\
    &=\frac{1}{N-1}+\frac{N(N-2)}{2(N-1)}(2-2^{1-r}),\\
     \rho^*_{total}(\mathbb{C}_N^{(r)})\leq \rho_{total}(\mathbb{C}_N^{(r)})&=\frac{1}{R_{FR}(\mathbb{C}_N^{(r)})}-1\\
     &= \frac{1}{N-1}+ \frac{N^2-N-1}{2(N-1)}(2-2^{1-r}).
\end{align}

\begin{Proof}
The algorithm builds on the multigraph scheme, by applying the PIR scheme across $r$ stages. Similar to the scheme on the respective simple graph $G$, the multigraph PIR scheme is repeated $x$ times by masking the answers. Also, $y$ symbols of $\cR$ are downloaded from each server. The subtle difference is that, beyond the first stage, symbols of $\cR$ are assigned to message sums. For example, compare Table~\ref{tab:answers fully rep multipath} for $\mathbb{P}_3^{(2)}$ with Table~\ref{tab:spir_p3_fully_rep} for $\mathbb{P}_3$. The steps of the algorithm are described below.

\begin{table}[ht]
    \centering
    \begin{tabular}{|c|c|c|c|}
    \hline
    & database 1 & database 2 & database 3\\
    \hline
   &  $s_1$ & $s_2$ & $s_8$\\
   \hline
   rep. 1 & $a_1+s_2$  & $a_2+b_2+s_1+s_3$ & $b_2+s_3$ \\
    & $c_1+s_4$ & $c_2+d_2+s_5+s_6$ & $d_2+s_6$\\
    & $a_3+c_2+s_5$ & $a_4+b_4+c_1+d_4+s_4+s_7$ & $b_4+d_4+s_7$\\
    \hline
    rep. 2 & $a_5+s_8$  & $a_6+b_8+s_8+s_9$ & $b_6+s_9$ \\
    & $c_5+s_{10}$ & $c_6+d_6+s_{11}+s_{12}$ & $d_6+s_{12}$\\
   & $a_7+c_6+s_{11}$ & $a_8+b_8+c_5+d_8+s_{10}+s_{13}$ & $b_8+d_8+s_{13}$\\
    \hline
    \end{tabular}
    \caption{Answer table for the SPIR scheme on $\mathbb{P}_3^{(2)}$ when $\theta=1$.}
    \label{tab:answers fully rep multipath}
\end{table}

\textit{Index Permutation:} For the SPIR scheme, let $L=2^{r-1}L'x$ corresponding to the $x$ repetitions of the multigraph scheme $T^{(r)}$. Also let $\cR$ constitute $H(\cR)=Ny+\frac{xL'}{2}\left((2^r-1)K'-1\right)$ i.i.d.~symbols, chosen uniformly at random by the servers from $\mathbb{F}_q$. The user applies private and independent permutations to the $L$ symbols of every message and to the $H(\cR)$ randomness symbols.

\textit{Query Generation:} Throughout the scheme, the user follows the query structure of the PIR scheme $T^{(r)}$ on the multigraph $G^{(r)}$, with $K=rK'$ messages. Similar to the algorithm for simple graphs, the said queries are repeated $x$ times. In each repetition, the user queries for $L=2^{r-1}L'$ new symbols of $W_{k,\tau}$. The user assigns common randomness symbols to the queries as follows.

\textit{Common Randomness Assignment:} As mentioned in Section~\ref{sec:achievable gr multi srp}, each repetition of $T^{(r)}$, proceeds in $r$ stages. Consider the first stage of the PIR scheme, where the messages $W_{k,t}$, for $t\in [r]$ are retrieved. Corresponding to the set of queries that retrieve $W_{k,\tau}$, the randomness assignment is identical to that of the simple graph algorithm (Section~\ref{subsec:achievable_algo}). In Table~\ref{tab:answers fully rep multipath}, this explains the assignment of $s_1,s_2,s_3,s_8$ and $s_9$. Additionally, corresponding to the queries that retrieve $W_{k,t},t\neq \tau$, a new randomness symbol is assigned for each distinct message symbol that is queried in the scheme. This explains the assignment of $s_4,s_5,s_6,s_{10},s_{11}$ and $s_{12}$. Because of SRP, this amounts to assigning 
\begin{align}
    \text{stage } 1: \underbrace{Ny+(K'-1)\frac{xL'}{2}}_{\text{assigned to queries for }W_{k,\tau}}+(r-1)
    \underbrace{\left(xL'+(K'-1)\frac{xL'}{2}\right)}_{\text{assigned to queries for  }W_{k,t}, t\neq \tau}
\end{align}
randomness symbols. In all subsequent stages of $T^{(r)}$, $s\in [2:r]$, the PIR scheme $T$ is applied on all $s$-sums of messages \cite{our_journal2025}. Correspondingly, in the SPIR scheme, we assign a unique randomness symbol of $\cR$ to each such $s$-sum of symbols. Thus, for each $A\subseteq[r], |A|=s$, and $s\in [2:r-1]$, we assign $xL'+(K'-1)\frac{xL'}{2}$ new randomness symbols to the queries that retrieve those $s$-sums that act as interference for the next stage, i.e., sums that do not involve symbols of $W_{k,\tau}$. On the other hand, for each $A\subseteq [r]$ containing $\tau$, the queries retrieve $s$-sums involving $W_{k,\tau}$. For each such query, we assign  $(K'-1)\frac{xL'}{2}$ new randomness symbols, corresponding to the $K'-1$ interferences in the retrieval. The remaining queries involve the symbols queried in the previous stage as $(s-1)$-sums, and the corresponding randomness symbols are reused. In the example, $s_7$ and $s_{13}$ are the new randomness symbols assigned to $b_4+d_4$ and $b_8+d_8$, respectively in stage 2, while $c_1+s_4, c_2+s_5$, $c_5+{s_{10}}$ and $c_6+s_{11}$ appear in both stages. In stage $s\in [2:r]$, the number of randomness symbols assigned is 
\begin{align}
    \text{stage } s: \binom{r-1}{s-1}\underbrace{(K'-1)\frac{xL'}{2}}_{\text{assigned to queries for }\sum_{m\in A:\tau\in A} W_{k,m}}+\binom{r-1}{s}\underbrace{\left(xL'+(K'-1)\frac{xL'}{2}\right)}_{\text{assigned to queries for }\sum_{m\in A:\tau\notin A}W_{k,m}}
\end{align}
across all repetitions. This justifies the size of $\cR$ to be
\begin{align}
    H(\cR) &= Ny+\sum_{s=1}^r\binom{r-1}{s-1}(K'-1)\frac{xL'}{2}+\sum_{s=1}^{r-1}\left(\binom{r-1}{s}\Big(xL'+(K'-1)\frac{xL'}{2}\Big)\right)\\
    &= Ny+\frac{xL'}{2}\left((2^r-1)K'-1\right).
\end{align}

\textit{Answer Formation:} Similar to the scheme of simple graphs, the user downloads $y$ randomness symbols from each server, followed by $x$ repetitions of PIR answers. To each answer, the servers add the assigned common randomness symbols. That is, an answer in stage $s$ in the form of an $st$-sum (where $t\geq 1$), is padded with $t$ symbols from $\cR$.

This completes the algorithm. The total number of downloaded symbols corresponds to $Ny$ randomness symbols, added to $x$ repetitions of the PIR scheme on $G^{(r)}$, to retrieve the $L$ symbols of $W_{k,\tau}$. Upon simplification, we get the rate expression \eqref{eq:ach rate multigraph fr}.

Reliability follows from the reliability of the underlying PIR scheme $T^{(r)}$, and the downloaded common randomness that mask $W_{k,\tau}$ in stage 1 across repetitions. 
The scheme supports user privacy, because the PIR scheme $T^{(r)}$ ensures that the queried message symbols are independent of $\theta$. The symbols of $\cR$ that are directly downloaded, and those added to the associated message sums appear to be chosen uniformly at random from $\cR$. This is due to the privately chosen permutations on the symbols of $\cR$.
To preserve database privacy, the answers are constructed such that every interference is one-time padded with a distinct symbol of $\cR$.
\end{Proof}

\begin{theorem}\label{thm:upperbnd_capacity fr multi}
    The SPIR capacity of $r$-multigraph extension of path and cyclic graphs are upper bounded as 
    \begin{align}
      \mathscr{C}_{FR}(\mathbb{P}_N^{(r)})&\leq \frac{2}{N(2-2^{1-r})+\frac{N}{N-1}2^{1-r} -(2-\frac{N}{N-1})(2-2^{1-r})},\\
        \mathscr{C}_{FR}(\mathbb{C}_N^{(r)})&\leq \frac{2}{(N+1)(2-2^{1-r})+\frac{N}{N-1}2^{1-r}-(2-2^{1-r})}.
    \end{align}
    The corresponding minimum total randomness ratios are lower bounded as
    \begin{align}
        \rho^*_{total}(\mathbb{P}_N^{(r)})&\geq \frac{1}{N-1}+\frac{N(N-2)}{2(N-1)}(2-2^{1-r})-\frac{3}{2}(2-2^{1-r}),\\
         \rho^*_{total}(\mathbb{C}_N^{(r)})&\geq \frac{1}{N-1}+\frac{N^2-N-1}{2(N-1)}(2-2^{1-r})-\frac{1}{2}(2-2^{1-r}).
    \end{align}
\end{theorem}

\begin{Proof}
The only modification in the converse proof, compared to the simple graph case comes from the inclusion of $(r-1)$ additional messages $\cw_{i,j}\setminus \{W_{k,\tau}\}$, replicated on the server pair $\{i,j\}$. We quantify this through the minimum sum of entropy of answers from servers $i$ and $j$, given $\cw\setminus \cw_{i,j}$, randomness and queries, when $W_{k,\tau}$ is the desired message. This is stated in Lemma~\ref{lem:answer pair servers multigraph}, which is the multigraph-version of Lemma~\ref{sum of answer entropies}. To prove this, we need the following SPIR extension of the interference bound lemma in \cite[Lemma 4.13]{our_journal2025}. 

\begin{lemma}\label{lem:interference in multi-edge}
    For $t\in [r-1]$ and any message $W_k$ replicated on servers $i$ and $j$, we have
    \begin{align}
        I(\cw_{i,j}\setminus W_{k,[t]};&A_i^{[k,t]},A_j^{[k,t]}|\cw\setminus \cw_{i,j},W_{k,[t]},\cR,\cq)\notag \\
        &\geq \frac{L}{2}+ \frac{1}{2}I(\cw_{i,j}\setminus W_{k,[t+1]};A_i^{[k,t+1]},A_j^{[k,t+1]}|\cw\setminus \cw_{i,j},W_{k,[t+1]},\cR,\cq). 
    \end{align}
\end{lemma}

\begin{Proof} 
\begin{align}
    2I(&\cw_{i,j}\setminus W_{k,[t]};A_i^{[k,t]},A_j^{[k,t]}|\cw\setminus \cw_{i,j},W_{k,[t]},\cR,\cq)\notag\\
    \geq& I(\cw_{i,j}\setminus W_{k,[t]};A_i^{[k,t]}|\cw\setminus \cw_{i,j},W_{k,[t]},\cR,\cq)\notag\\
    &+ I(\cw_{i,j}\setminus W_{k,[t]};A_j^{[k,t]}|\cw\setminus \cw_{i,j},W_{k,[t]},\cR,\cq)\\
    =& H(A_i^{[k,t]}|\cw\setminus \cw_{i,j},W_{k,[t]},\cR,\cq)+ H(A_j^{[k,t]}|\cw\setminus \cw_{i,j},W_{k,[t]},\cR,\cq)\label{eq:answers i and j deterministic}\\
    =& H(A_i^{[k,t+1]}|\cw\setminus \cw_{i,j},W_{k,[t]},\cR,\cq)+ H(A_j^{[k,t+1]}|\cw\setminus \cw_{i,j},W_{k,[t]},\cR,\cq\label{eq: holds by lemma user privacy}\\
    \geq& H(A_i^{[k,t+1]}, A_j^{[k,t+1]}|\cw\setminus \cw_{i,j},W_{k,[t]},\cR,\cq)\\
    =&H(W_{k,t+1}|\cw\setminus \cw_{i,j},W_{k,[t]},\cR,\cq)+H(A_i^{[k,t+1]},A_j^{[t+1]}|\cw\setminus \cw_{i,j},W_{k,[t+1]},\cR,\cq)\label{eq:since W_k,t+1 is decodable}\\
    =&L + H(A_i^{[k,t+1]},A_j^{[k,t+1]}|\cw\setminus \cw_{i,j},W_{k,[t+1]},\cR,\cq) - H(A_i^{[k,t+1]},A_j^{[k,t+1]}|\cw,\cR,\cq)\\
     =& L+I(\cw_{i,j}\setminus W_{k,[t+1]};A_i^{[k,t+1]},A_j^{[k,t+1]}|\cw\setminus \cw_{i,j},W_{k,[t+1]},\cR,\cq), 
\end{align} 
where \eqref{eq: holds by lemma user privacy} holds by Lemma~\ref{lem: conseq of user priv},  and \eqref{eq:since W_k,t+1 is decodable} holds because $W_{k,t+1}$ is decodable from $A_{i}^{[k,t+1]}$ and $A_{j}^{[k,t+1]}$, given $\cw\setminus \cw_{i,j},\cR, \cq$. 
\end{Proof}

\begin{lemma}\label{lem:answer pair servers multigraph}
    For $k\in [K']$, let $\cw_{i,j}=\{W_{k,t}, t\in [r]\}$ be replicated on servers $i$ and $j$. Then, for any $\tau\in [r]$,
    \begin{align}
        H(A_i^{[k,\tau]}&|\cw\setminus \cw_{i,j},\cR,\cq)+H(A_j^{[k,\tau]}|\cw\setminus\cw_{i,j},\cR,\cq)\notag\\
        &\geq L\left(1+\frac{1}{2}+\frac{1}{2^2}+\ldots+\frac{1}{2^{r-1}}\right)=L(2-2^{1-r}).
    \end{align}
\end{lemma}

\begin{Proof}
This holds since, $ H(A_i^{[k,\tau]}|\cw\setminus \cw_{i,j},\cR,\cq)+H(A_j^{[k,\tau]}|\cw\setminus\cw_{i,j},\cR,\cq)\geq H(A_i^{[k,\tau]},A_j^{[k,\tau]}|\cw\setminus \cw_{i,j},\cR,\cq)= I(A_i^{[k,\tau]},A_j^{[k,\tau]};\cw_{i,j}|\cw\setminus \cw_{i,j},\cR,\cq)$. To bound this, we apply Lemma~\ref{lem:interference in multi-edge} successively to $t=1,2,\ldots,r$.
\end{Proof}

Next, we state the multigraph-versions of Lemmas \ref{lem:ans bound path} and \ref{lem:answer bound cycle}, which can be proved similarly as the simple graph versions.

\begin{lemma}\label{lem:multigraph answer bound}
    For any $\tau\in [r]$, the following bounds hold: 
    \begin{itemize}
    \item For path multigraphss $\mathbb{P}_N^{(r)}$, 
        \begin{align}
            H(A_{2:N}^{[1,\tau]}|A_1^{[1,\tau]},W_1,\cq)&\geq \frac{(N-2)(2-2^{1-r})L}{2}, \\
            H(A_{1:N-1}^{[N-1,\tau]}|A_{N}^{[N-1,\tau]},W_{N-1},\cq) & \geq \frac{(N-2)(2-2^{1-r})L}{2}, \\
            H(A_{1:N\setminus \{n\}}^{[k,\tau]}|A_n^{[k,\tau]},W_n,\cq)&\geq \frac{(N-3)(2-2^{1-r})L}{2}, k\in \{n-1,n\}, n\in \{2,\ldots,N-2\}. \label{eq:answers of non-edge servers multi}
        \end{align}
    \item For cyclic multigraphs $\mathbb{C}_N^{(r)}$,
        \begin{align}
            H(A_{1:N\setminus \{n\}}^{[k,\tau]}|A_n^{[k,\tau]},W_n,\cq)&\geq \frac{(N-2)(2-2^{1-r})L}{2}, \, k\in \{n-1,n\}.
        \end{align}
    \end{itemize}
\end{lemma}

For any $\tau\in [r]$, similar steps give us 
\begin{align}
    L&\leq H(A_{1:N}^{[k,\tau]}|\cq) - H(A_n^{[k,\tau]}|\cq)-H(A_{1:N\setminus \{n\}}^{[k,\tau]}|A_n^{[k,\tau]},W_{k,\tau},\cq). \label{eq:first step upper bound}
\end{align}
From here, invoking Lemma~\ref{lem:multigraph answer bound} on \eqref{eq:first step upper bound}, we obtain the required capacity bounds. Moreover, using Lemma~\ref{lem:randomness size fr} and the same steps as the simple graph, we derive the total randomness ratio bounds. This concludes the proof of Theorem~\ref{thm:upperbnd_capacity fr multi}.  
\end{Proof}

\section{Conclusion}\label{sec:conclude_spir}
In this work, we initiated the study of SPIR on graph-based replicated systems. Our work is motivated by the partially-replicated nature in which sensitive messages are stored, or made accessible to users, across servers. This makes database privacy to be crucial for PIR. We focused on 2-replicated database systems where the common randomness to enable SPIR is either graph-replicated or fully-replicated at the servers. The database privacy requirements enforced by both settings are distinct; with the former setting being more restrictive. 

For the first setting, we settled the SPIR capacity for regular and path (multi)graph families, and is open for general graphs. An interesting question to explore is whether there are graphs for which the capacity is strictly greater than $\frac{1}{N}$. Moreover, the capacity is independent of the number of messages $r$ replicated on each server pair. 

For the second setting, we solved the SPIR problem  exactly for $\mathbb{P}_3$. We established bounds on the capacity for path and cyclic (multi)graphs, which revealed connections to the respective PIR capacities. 

There is room to sharpen our bounds and derive non-trivial upper bounds on the SPIR capacity of other graph families, e.g., complete and star graphs. In this work, we explored the impact of two extreme cases of common randomness replication in graph-based replicated systems. It would be interesting to explore the interactions between the replication of server-side common randomness and message replication for the feasibility of SPIR. 

\bibliography{references}
\bibliographystyle{unsrt}
\end{document}